\documentclass[preprint, showpacs, showkeys, preprintnumbers, amsmath,amssymb]{revtex4}

\usepackage{graphicx}  
\usepackage{dcolumn}   
\usepackage{bm}        

\usepackage{amssymb, amsmath}
\usepackage{mathrsfs}			

\newcommand{\FP}{{\mathscr P}}		
\newcommand{\mapping}{{\Phi}}		
\newcommand{\inmapping}{{\Phi^{-1}}}	

 \newcommand{\ve}[1]{{\bf #1}}
 \newcommand{\vve}[1]{{\bf #1}}
 \newcommand{\veg}[1]{{\boldsymbol {#1}}}	
 \newcommand{\vveg}[1]{{\boldsymbol {#1}}}	

 \newcommand{\sgn} {{\rm sgn}}
 \newcommand{\const} {{\rm const}}

\newcommand{\Henon} {{H$\rm\acute e$non}}

\newcommand{\matthree}[9]{\bracket{\begin{array}{ccc}
		#1	&#2	&#3	\\
		#4	&#5	&#6	\\
		#7	&#8	&#9
		\end{array}}}

\newcommand{\mat}[4]{\bracket{\begin{array}{cc}
		#1	&#2\\
		#3	&#4	   	
		\end{array}}}

\newcommand{\D}{\partial}
\newcommand{\DD}[2]{\frac {\D #1}{\D #2}}
\newcommand{\Dt}[1]{\frac {\D #1} {\D t}}
\newcommand{\Dx}[1]{\frac {\D #1} {\D x}}

\newcommand{\dt}[1]{\frac {d #1} {d t}}
\newcommand{\delt}{{\Delta t}}
\newcommand{\hot}{o(\delt)}
\newcommand{\delw}{{\Delta\ve w}}

\newcommand{\DI}[1]{\frac {\D #1} {\D x_1}}	

\newcommand{\DIII}[1]{\frac {\D #1} {\D x_3}}
\newcommand{\Dn}[1]{\frac {\D #1} {\D x_n}}

\newcommand{\Di}[1]{\frac {\D #1} {\D x_i}}

\newcommand{\DIDI}[1]{\frac {\D^2 #1} {\D x_1^2}}

\newcommand{\DyI}[1]{\frac {\D #1} {\D y_1}}	
\newcommand{\DyII}[1]{\frac {\D #1} {\D y_2}}

\newcommand{\Dyn}[1]{\frac {\D #1} {\D y_n}}
\newcommand{\Dyi}[1]{\frac {\D #1} {\D y_i}}
\newcommand{\Dyj}[1]{\frac {\D #1} {\D y_j}}

\newcommand{\DyiDyj}[1]{\frac {\D^2 #1} {\D y_i \D y_j}}
\newcommand{\DyIDyI}[1]{\frac {\D^2 #1} {\D y_1^2}}

\newcommand{\Jacobian}[2]{{\left[\frac{\D(#1)}{\D(#2)}\right]}}

\newcommand{\excl}[1]{{\backslash \hspace{-0.3em} #1}}

\newcommand{\abs}[1]{\left|#1\right|}
\newcommand{\bracket}[1]{\left[#1\right]}
\newcommand{\cbrace}[1]{\left\{#1\right\}}
\newcommand{\parenth}[1]{\left(#1\right)}
\newcommand{\II}{I\!\!I}
\DeclareSymbolFont{AMSb}{U}{msb}{m}{n}
\DeclareMathSymbol{\R}{\mathbin}{AMSb}{"52}

\newcommand{\pf} {\noindent {\bf Proof.\ }}
\newcommand{\qed} {$\square$}

\newtheorem{thm}{Theorem}[section]
\newtheorem{corollary}{Corollary}[section]

\newtheorem{prop}{Proposition}[section]

\begin{document}

 \title{\bf Information flow and causality\\
	 as rigorous notions {\it ab initio}}

\author{X. San Liang}
\email{sanliang@courant.nyu.edu, URL: http://www.ncoads.org.}
\affiliation{Nanjing Institute of Meteorology, Nanjing 210044, and\\
China Institute for Advanced Study, Beijing 100081, China}

         \date{\today}

\begin{abstract}
Information flow (or information transfer as may be called)
the widely applicable general physics notion
can be rigorously derived from first principles, 
rather than axiomatically proposed as an ansatz.
Its logical association with causality and, 
particularly, the most stringent one-way
causality, if existing, is firmly substantiated and
stated as a fact in proved theorems. 
Established in this study are the information flows among
the components of time-discrete mappings and time-continuous dynamical
systems, both deterministic and stochastic.
They have been obtained explicitly in closed form, and all possess the
property of causality, which reads: if a component, say $x_i$, has
an evolutionary law independent of $x_j$, then the information flow 
from $x_j$ to $x_i$ vanishes. 
These results have been put to applications with benchmark systems, 
such as the Kaplan-Yorke map, the R\"ossler system,
the baker transformation, 
the H\'enon map, 
and a stochastic potential flow.
Besides recovering the properties as expected from the respective
systems, some of the applications show that the information flow structure 
underlying a complex trajectory pattern could be tractable.
For linear systems, the resulting remarkably concise formula
asserts analytically that causation implies correlation, 
while correlation does not imply causation, resolving unambiguously
the long-standing debate over causation versus correlation.

\end{abstract}



\maketitle


%





\section{Introduction}

{I}{n}formation flow, or information transfer as it may be
referred to in the literature, has been realized as a fundamental notion in
general physics. Though literally one may associate it with communication, its
importance lies far beyond 
in that it implies causation\cite{Schindler}-\cite{Schreiber},
uncertainty propagation\cite{Liang11}, 
predictability transfer\cite{Kleeman08}, etc. 
In fact, it is the recognition of its causality association that has 
attracted enormous interest from a wide variety
of disciplines, particularly in 
neuroscience\cite{Pereda}-\cite{Wu}, 
finance\cite{Kantz}-\cite{Lee},
climate science\cite{Wang}-\cite{Runge},
turbulence research\cite{Tissot}-\cite{Materassi}, 
network dynamics\cite{RSun}-\cite{Timme}, 
and dynamical systems particular in the field of
synchronization\cite{Pikovsky}-\cite{Pethel05}.
This recognition has been further substantiated by the finding that
transfer entropy\cite{Schreiber} and Granger causality\cite{Granger69} 
are equivalent (up to a factor 2)\cite{Barnett}.

Historically, many information theoretic quantities have been proposed to
measure information flow, including time-delayed mutual
information\cite{Swinney},
transfer entropy\cite{Schreiber}, 
momentary information transfer\cite{Runge}, 
causation entropy\cite{JieSun}, to name a few.
Among these most notably is transfer entropy,
which has spawned many varieties 
in its family, e.g., \cite{Kantz}, \cite{Duan}, \cite{Staniek},
and has been widely applied in different disciplines.

A fundamental question to ask is whether information flow needs to be
axiomatically proposed as an ansatz (as above), 
or it can be derived from the first principles
in information theory. Naturally, one would like to minimize 
or avoid the use of axioms in introducing new concepts in order to have 
the material more coherent within the field to which it belongs.
In physics, ``flow'' or ``transfer''
does have definite meaning, albeit the meaning may differ depending on 
the context. One then naturally expects that the concept be rigorized.
Indeed, as we will see soon, at least within the framework of dynamical
systems, information flow/transfer can be rigorously derived from, 
rather than empirically or axiomatically proposed with, Shannon entropy.

Another impetus regards the inference of causality. As mentioned in the
beginning, information flow arouses enormous interest in a wide range of
fields not because of its original meaning in communication but because of
its logical implication of causation. Whether the cause-effect relation 
underlying a system can be faithfully revealed is, therefore, the touchstone
for a formalism of information flow. That is to say, information flow
should be formulated with causality naturally embedded; it should, in
particular, accurately reproduce a one-way causality (if existing), 
which is unambiguously equal to zero on one side. 
In this light, the widely used formalism
namely transfer entropy is, unfortunately, not as satisfactory one expects. 
This has even led to discussions on whether the two notions, namely,
information flow and causality , should be differentiated (e.g.,
\cite{Lizier}).
Since it is established that Granger causality and transfer entropy are 
equivalent, one may first look at the problems from the former.
Now it is well known that spurious Granger causality may arise due to
unobserved variables that influence the system dynamics (a problem
identified by Granger himself)\cite{Granger80}, due to low resolution in
time\cite{Sims}\cite{Smirnov12}, and due to observational
noise\cite{Nalatore}. Besides, Granger explicitly excludes deterministic
systems in establishing the causality formalism, a case that for sure is
important in realistic problems.
For transfer entropy, the issue has just been systematically
examined\cite{Smirnov13}.
Aside from the failure in recovering the many preset one-way causalities, 
evidence has shown that sometimes it may even give qualitatively wrong
results; see \cite{Smirnov13} and \cite{Hahs} for such examples. 

Realizing the limitation of transfer entropy, 
different alternatives have been proposed;
the above momentary information transfer is one of these proposals.
The purpose of this study is, instead of just remedying the deficiencies
of the existing formalisms, to put information flow the fundamental
physical notion on a rigorous footing so that it is universally applicable.
The stringent one-way causality requirement will not be just verified with
certain given examples, but rigorously proved as theorems.

With this faith, recently Liang and Kleeman (2005)\cite{LK05} take the 
initiative to study the problem with dynamical systems. 
In this framework, the information source and recipient are abstracted as the
system components, and hence the problem is converted into the information
flow or information transfer between dynamical system components. 
The basic idea can be best illustrated with a deterministic system of two
components, say, $x_1$ and $x_2$:
	\begin{eqnarray}
	\dt{x_1} = F_1(x_1,x_2,t), \\
	\dt{x_2} = F_2(x_1,x_2,t),
	\end{eqnarray} 
where we follow the convention in physics and do not distinguish
random and deterministic variables, which should be clear in the context.
Now what we are to consider are the time evolutions of the 
marginal entropies of $x_1$ and $x_2$, denoted respectively 
as $H_1$ and $H_2$. Look at $x_1$, its marginal entropy evolution
may be due to $x_1$ itself or subject to the influence of $x_2$.
This partitions the mechanisms that cause $H_1$ to grow into two exclusive
parts. That is to say, if we write the contribution from the former mechanism 
as $dH^*/dt$ and that from the latter as $T_{2\to1}$, 
	\begin{eqnarray}	\label{eq:LK_decomposition}
	\dt {H_1} = \dt {H_1^*} + T_{2\to1}.
	\end{eqnarray}
This $T_{2\to1}$ is the very time rate of information flowing 
from $x_2$ to $x_1$. 
We remark that this setting is rather generic, except for the requirement
of differentiability for the vector field $\ve F = (F_1,F_2)^T$.
In particular, the input-output communication problem can be cast within the
framework by letting, for example, $F_2 = F_2(x_1,t)$, $F_1=F_1((x_1,t)$, 
where $x_1$ is the input/drive and $x_2$ the output/consequence, and the
channel is represented by $F_2$.

From the above argument, the evaluation of the information flow 
$T_{2\to1}$ may be fulfilled through evaluating $dH_1^*/dt$.
This is because that, when a dynamical system is given, the density
evolution is known through the corresponding Liouville equation, and,
accordingly, $dH_1/dt$ can be obtained. 
In \cite{LK05}, Liang and Kleeman prove that the joint entropy of
$(x_1,x_2)$ follow a very concise law
	\begin{eqnarray}
	\dt H = E(\nabla\cdot\ve F),
	\end{eqnarray}
where $E$ is the operator of mathematical expectation. They then argue that
	\begin{eqnarray}	\label{eq:phys_argu}
	\dt {H_1^*} = E\parenth{\DI {F_1}},
	\end{eqnarray}
and hence obtain the time rate of information flowing from 
$x_2$ to $x_1$ 
	\begin{eqnarray}	\label{eq:T21}
	T_{2\to1} = \dt {H_1} - \dt {H_1^*} 
		  = -E\parenth{\frac 1 {\rho_1} \DI {F_1\rho_1}},
	\end{eqnarray}
where $\rho_1$ is the marginal probability density function of $x_1$.
The thus-obtained  information flow is asymmetric 
between $x_1$ and $x_2$; moreover, it possesses a 
{\it property of causality}, which reads,
if the evolution of $x_1$ does not depend on $x_2$, 
then $T_{2\to 1} = 0$. 


%

The above result is later on proved\cite{LK07a}\cite{LK07b}. It is
remarkable in that the stringent one-way causality in a system, 
if existing, can be stated as a proven theorem, rather than a fact for a
formalism to verify; see \cite{Liang13} for a review. This result, however,
is only for systems of dimension 2 (2D). For systems with 
many components, it does not work any more. 
We have endeavored to extend it to more general situations and do have
obtained results for deterministic systems of arbitrary dimensionality
which possess the property of causality. But, as we will see in the 
following section, the extension relies on an assumption that is, again,
axiomatically proposed. This makes the resulting formalism not one fully 
derived from first principles, and as we realize later on, it does not work for
multidimensional stochastic systems. This line of work, though with a 
promising start, is stuck at this point.

In this study, we will show that the assumption can be completely removed.
In a unified approach, the notion of information flow can be rigorously
derived for both deterministic and stochastic systems of arbitrary
dimensionality.  In the following, we first briefly set up the framework, 
and show where the snag lies in the above approach. The solution is then
presented, and applied to derive the information flows for deterministic
mappings (section~\ref{sect:det_map}), continuous-time deterministic
systems (section~\ref{sect:det_flow}), stochastic mappings
(section~\ref{sect:stoch_map}), and continuous-time stochastic systems 
(section~\ref{sect:stoch_flow}). For the purpose of demonstration, each
section contains one ore more applications. As an important particular
case, we specialize to do the derivation for linear systems, and the
material is presented in section~\ref{sect:linear}. This study is
summarized in section~\ref{sect:summary}.

\section{The snag that stuck the Liang-Kleeman formalism}

The success of the Liang-Kleeman formalism is remarkable. 
It is, however, only for 2D dynamical systems.
When the dimensionality exceeds 2, the resulting quantity,
namely, (\ref{eq:T21}),
is not the information transfer from $x_2$ to $x_1$, but the
cumulant transfer to $x_1$ from all other components $x_2$, $x_3$,..., $x_n$.
In this sense, the use of (\ref{eq:T21}) is rather limited.

In order to extend the formalism to systems of higher dimensionality,
Liang and Kleeman\cite{LK07a}\cite{LK07b} re-interpret the term
$dH_1^*/dt$ in the above decomposition (\ref{eq:LK_decomposition}), 
for a 2D system,
as the evolution of $H_1$ with the effect of $x_2$ excluded. 
More specifically, it is the evolution of $H_1$ with $x_2$ instantaneously
frozen as a parameter at time $t$. To avoid confusing 
with $dH_1^*/dt$, denote it as $dH_{1\excl2}/dt$, where 
the subscript $\excl2$ signifies that $x_2$ is frozen, or that its effect
is removed. With this, the disjoint decomposition (\ref{eq:LK_decomposition})
is re-stated as
	\begin{eqnarray}	\label{eq:LK_decomposition2}
	\dt {H_1} = \dt {H_{1\excl2}} + T_{2\to1}.
	\end{eqnarray}
Note this decomposition, albeit seemingly with only a change of symbol,
is actually fundamentally different from (\ref{eq:LK_decomposition}) in
physical meaning; it now holds for systems of arbitrary dimensionality.
The information flow is, therefore, 
	\begin{eqnarray}	\label{eq:T21a}
	T_{2\to1} = \dt {H_1} - \dt {H_{1\excl2}}.
	\end{eqnarray}

Of course, the key is how to find $\dt {H_{1\excl2}}$. In \cite{LK07a}
and \cite{LK07b}, Liang and Kleeman start with discrete mappings, and then
take the limit as the time stepsize goes to zero. To illustrate,
let $\mapping: \R^n \to \R^n$ be a mapping taking $\ve x(\tau)$ to 
$\ve x(\tau+1)$, from time step $\tau$ to $\tau+1$.
Correspondingly there is another mapping $\FP: L^1(\R^n) \to L^1(\R^n)$
that steers its density $\rho$ forward. This mapping is called 
a Frobenius-Perron operator; we will refer it to F-P operator henceforth.
Loosely speaking, $\FP$ is, for any $\omega\subset\R^n$, such
that\cite{Lasota} 
	\begin{eqnarray}	\label{eq:FPdefinition}
	\int_\omega \FP\rho(\ve x) d\ve x = 
	\int_{\mapping^{-1}(\omega)}\rho(\ve x) d\ve x.
	\end{eqnarray} 
When the sample space is in a Cartesian product form, 
as is in this case ($\R^n$), the operator can be evaluated. Let 
$\ve a = (a_1,a_2,...,a_n)$ be some constant point, 
and $\omega = [a_1,x_1] \times [a_2,x_2] \times ... \times[a_n,x_n]$.  
It has be established that (e.g., \cite{Lasota})
	\begin{eqnarray}	\label{eq:FPcompute}
	\FP\rho(\ve x) = \frac {\D^n} {\D x_n ... \D x_2\D x_1}
	\int_{\inmapping(\omega)}
	\rho(\xi_1,\xi_2,...,\xi_n) d\xi_1d\xi_2...d\xi_n.
	\end{eqnarray}
For convenience, $\ve a$ is usually taken to be the origin. Furthermore, if
$\mapping$ is nonsingular and invertible, then $\FP$ can be explicitly
written out
	\begin{eqnarray}	\label{eq:FP_explicit}
	\FP\rho(\ve x) = \rho\bracket{\inmapping(\ve x)} \cdot |J^{-1}|
	\end{eqnarray}
where $J$ is the Jacobian of $\mapping$.
 
As the F-P operator carries $\rho$ forth from time step $\tau$ to $\tau+1$,
accordingly the entropies $H$, $H_1$, and $H_2$ are also steered forward.
On $[\tau,\tau+1]$, let $H_1$ be incremented by $\Delta H_1$. 
By the foregoing argument, 
the evolution of $H_1$ can be decomposed into two exclusive
parts, namely, the information flow from $x_2$, $T_{2\to1}$, 
and the evolution with the effect of $x_2$ excluded, $\Delta H_{1\excl2}$.  
Hence, for discrete mappings, we have the following
counterpart of (\ref{eq:T21a}):
	\begin{equation}	\label{eq:T21_dis}
        T_{2\to1} = \Delta {H_1} - \Delta {H_{1\excl2}}.
	\end{equation}

Liang and Kleeman derive the information flow for the continuous system
from the discrete mapping. So the whole procedure relies on how 
  	$$\Delta H_{1\excl2} = H_{1\excl2}(\tau+1) - H_1(\tau)$$ 
is evaluated, or, more specifically, how $H_{1\excl2}(\tau+1)$ 
is evaluated (since $H_1(\tau)$ is known). 
To see where lies its difficulty, first notice that
	\begin{eqnarray*}
	H_1(\tau+1) = - \int_\R (\FP\rho)_1(x_1) \log(\FP\rho)_1(x_1) dx_1
	\end{eqnarray*}
which is the mean of $-\log(\FP\rho)_1(x_1)$. Given $\mapping$, $\FP$ can be
found in the way as shown above, so $H_1(\tau+1)$ is known. For $H_{1\excl2}$,
however, things are much more difficult;
$-\log(\FP_\excl2\rho)_1(x_1)$ involves not only the random variable
$x_1(\tau+1)$, but also $x_2(\tau)$ (embedded in the subscript $\excl2$). 
What is the joint density of 
$(x_2(\tau), x_1(\tau+1))$? We do not know.
In \cite{LK07a} and \cite{LK07b}, an approximation was proposed, which
gives
	\begin{eqnarray*}
	H_{1\excl2}(\tau+1) = -\int_\Omega (\FP_{\excl2}\rho)_1 (y_1) 
	 \log(\FP_{\excl2}\rho)_1(y_1) \cdot \rho(x_2 | x_1, x_3,...,x_n)
	 \cdot \rho_{3...n}(x_3,...,x_n) \ dy_1 dx_2 dx_3 ... dx_n,
	\end{eqnarray*}
where $y_1$ is employed to signify $x_1(\tau+1)$  and the symbol $x_1$ 
is reserved
for $x_1(\tau)$.
This is a natural extension of what the authors use in the their original
study\cite{LK05} for 2D discrete mappings. A central
approximation is that they use 
	\begin{eqnarray*}
	\rho(x_2 | x_1, x_3, ..., x_n) \cdot
	(\FP_{\excl2}\rho)_1 (y_1) \rho_{3...n}(x_3,...,x_n) 
	\end{eqnarray*}
to represent the joint pdf of $y_1$ and $x_2$ (and $x_3,...x_n$)
(think about $\rho(x_2|x_1) \rho(x_1) = \rho(x_1,x_2)$). This is, however,
only an approximation, since we really don't know what the joint pdf of 
$(y_1,x_2)$ is. As we will see soon, though the resulting formalism verifies
$dH_{1\excl2}/dt = dH_1^*/dt$ for 2D systems and the zero-causality
property for one-way causal deterministic systems, 
when stochasticity gets in, the
causality property cannot be recovered this way.

\section{Deterministic mapping}	\label{sect:det_map}

\subsection{Derivation}

Fortunately, the issue that stuck the Liang-Kleeman formalism can be fixed;
we actually can get the entropy without appealing to the joint
probability density function of $(y_1,x_2)$, i.e., that of $(x_1(\tau+1),
x_2(\tau))$ as mentioned above.
Consider a mapping 
    \begin{eqnarray*}
      \mapping: \Omega \to \Omega, \qquad
      \ve x(\tau) \mapsto \ve x(\tau+1) = 
	\parenth{\mapping_1(\ve x), \mapping_2)\ve x), ..., \mapping_n(\ve x)},
    \end{eqnarray*}
where $\Omega$ is the sample space ($\R^n$ in particular).
Let $\psi: \Omega \to\Omega$ be an arbitrary differentiable function of $\ve x$.
We have the following theorem:
	\begin{thm}	\label{thm:mean_equality}
	\begin{eqnarray}	\label{eq:mean_equality}
	E\psi(\ve x(\tau+1)) = E\psi(\mapping(\ve x(\tau))).
	\end{eqnarray}
	\end{thm}
Remark~1: The expectation operator $E$ on the right hand side 
	applies to a function of 
	$\ve x(\tau)$; it is thence with respect to $\rho(\tau)$. 
	Differently, the left hand side $E$ is with respect to 
	$\rho(\tau+1) = \FP\rho$, where
	$\FP$ is the F-P operator as introduced in (\ref{eq:FPdefinition}).

Remark~2: This equality is important in that one actually can obtain the
expectation of $\psi(\ve x(\tau+1))$ without evaluating $\FP\rho$.

\pf
The following proof is in the framework of Riemann-Stieltjes integration. A
more general proof in terms of Lebesgue theory is also possible but is
unnecessary, since the functions and vector fields we are dealing with 
in this study are assumed to differentiable.

Let $\{\omega_1,\omega_2,...,\omega_n\}$ be a partitioning of the sample
space $\Omega$. The elements are mutually exclusive and 
$\Omega = \cup_{k=1}^n \omega_k$. To make it simple, assume that
these $\omega_k$'s have the same diameter (the maximal distance between any
two points in $\omega_k$).
For clarity, write $\ve x(\tau+1)$ as $\ve y$, while $\ve x$ is reserved
for $\ve x(\tau)$. Then
	\begin{eqnarray*}
        && E\psi(\ve x(\tau+1)) 
		= \int_\Omega \FP\rho(\ve y) \psi(\ve y) d\ve y	\\
	&& = \lim_{n\to\infty} \sum_{k=1}^n \int_{\omega_k} 
		\FP\rho(\ve y)\psi(\ve y) d\ve y
	   = \lim_{n\to\infty} \sum_{k=1}^n \psi(\ve y_k) \int_{\omega_k}
		\FP\rho(\ve y) d\ve y,
	\end{eqnarray*}
where $\ve y_k \in\omega_k$ is some point in $\omega_k$. 
The existence of the Riemann
integral $\int_\Omega \FP\rho(\ve y) \psi(\ve y) d\ve y$ assures that it
can be any point in $\omega_k$ as $n$ goes to infinity, 
while the resulting integral is the same.
Now by (\ref{eq:FPdefinition}), 
	\begin{eqnarray*}
	\int_{\omega_k}\FP\rho(\ve y) d\ve y
	= \int_{\inmapping(\omega_k)} \rho(\ve x) d\ve x.
	\end{eqnarray*}
So the above becomes
	\begin{eqnarray*}
        && E\psi(\ve x(\tau+1)) 
	   = \lim_{n\to\infty} \sum_{k=1}^n \psi(\ve y_k) \int_{\omega_k}
		\FP\rho(\ve y) d\ve y				\\
	&&  = \lim_{n\to\infty} \sum_{k=1}^n \psi(\ve y_k) 
		\int_{\inmapping(\omega_k)}\rho(\ve x) d\ve x  	\\
	&&  = \lim_{n\to\infty} \sum_{k=1}^n 
		\int_{\inmapping(\omega_k)}\rho(\ve x) 
		\psi(\mapping(\ve x)) d\ve x.
	\end{eqnarray*}
Notice, for $\mapping:\Omega\to\Omega$, $\Omega=\cup_k\omega_k$,
it must be that $\cup_k \inmapping(\omega_k) = \Omega$.
So the limit converges to 
	$\int_\Omega \rho(\ve x) \psi(\mapping(\ve x)) d\ve x$.
That is to say, $E\psi(\ve x(\tau+1)) = E\psi(\mapping(\ve x(\tau)))$.
\qed

The equality (\ref{eq:mean_equality}) actually can be utilized to 
derive the F-P operator. We look at the particular case when $\mapping$ is
invertible. By definition, Eq.~(\ref{eq:mean_equality}) means
	\begin{eqnarray*}
	\int_\Omega \psi(\ve x) \rho(\tau+1,\ve x) d\ve x
	= \int_\Omega \psi(\mapping(\ve x)) \rho(\tau,\ve x) d\ve x.
	\end{eqnarray*}
If $\mapping$ is invertible, the right hand side is
	$\int_\Omega \psi(\ve y) \cdot \rho\parenth{\tau, \inmapping(\ve y)}
	\abs{J^{-1}}\ d\ve y$
by transformation of variables.
Since $\psi$ is arbitrary, we have
	$$\FP\rho = \rho(\tau+1,\ve x) = \rho(\tau, \inmapping(\ve x)) 
		   \cdot \abs{J^{-1}},$$
which is precisely the Frobenius-Perron operator (\ref{eq:FP_explicit}).

The above equality provides us a convenient and accurate way to evaluate 
$H_1(\tau+1)$ and $H_1\excl2(\tau+1)$. Picking $\psi$ as $(\log\FP\rho)_1$ and 
$\log(\FP_\excl2\rho)_1$, we obtain, respectively, the following formulas:
	\begin{corollary}
	\begin{eqnarray}
	&&H_1(\tau+1) = - E\log(\FP\rho)_1(\mapping_1(\ve x)),\\
	&&H_{1\excl2}(\tau+1) = - E\log(\FP_\excl2\rho)_1(\mapping_1(\ve x)).
		\label{eq:H1no2_dis}
	\end{eqnarray} 
	\end{corollary}
In these formulas,
both the expectations are taken with respect to $\rho(x_1,x_2,...x_n)$,
i.e., the pdf at time step $\tau$. In (\ref{eq:H1no2_dis}), we do not need
to care about the joint pdf $\rho(y,x_2)$ any more. 
The information flow from $x_2$ to $x_1$ is, therefore,
	\begin{thm}
	\begin{eqnarray}	\label{eq:T21_det_dis}
	T_{2\to1} = E\log(\FP_\excl2\rho)_1(\mapping_1(\ve x))
		  - E\log(\FP\rho)_1(\mapping_1(\ve x)).
	\end{eqnarray}
	\end{thm}
\pf
	\begin{eqnarray*}
	T_{2\to1} = \Delta H_1 - \Delta H_{1\excl2}
	= (H_1(\tau+1) - H_1(\tau)) -
	  (H_{1\excl2}(\tau+1) - H_1(\tau)) 
	= H_1(\tau+1) - H_{1\excl2}(\tau+1).
	\end{eqnarray*}
Substitute into the above formulas for $H_1$ and $H_{1\excl2}$
and (\ref{eq:T21_det_dis}) follows.

Note that the evaluation of $\FP_\excl2\rho$ and $\FP\rho$ generally
depends on the system in question. But when $\mapping$ and
$\mapping_\excl2$ are invertible, the information flow can be found
explicitly in a closed form.

\subsection{Properties}

   \begin{thm}		\label{thm:2d_dis}
   For 2D systems, if $\mapping_1$ is invertible, then
	$$\Delta H_{1\excl2} = H_{1\excl2}(\tau+1) - H_1(\tau)
	 = E\log|J_1|.$$
   \end{thm}
Remark:
This is the analog of (\ref{eq:phys_argu}) for discrete-time 
systems\cite{LK05}. 

   \pf
   Let $\ve x(\tau+1) \equiv \ve y$. For a 2D system, and if $\mapping_1$ is
   invertible, we have
	$$(\FP_\excl2\rho)_1(\ve y) = \rho_1(\mapping_1^{-1}(y_1)) 
	   \cdot |J_1^{-1}|,$$
   which gives
	\begin{eqnarray*}
	H_{1\excl2} 
	&=& - E_x \log \bracket{\rho_1\parenth{\inmapping_1(y_1)} 
		\cdot |J_1^{-1}| }	\\
	&=& - E \log \bracket{\rho_1(x_1)) \cdot |J_1^{-1}|} \\
	&=& - E \log\rho_1(x_1) + E\log|J_1|.
	\end{eqnarray*}
To avoid confusion, we use $E_x$ to indicate that the expectation is
with respect to $x$ when mixed variables $x$ and $y$ appear simultaneously.
Note here $x_1 = \inmapping_1(y_1)$ since this is a 1D system after $x_2$ is
frozen. Thus
	$$\Delta H_{1\excl2} = E\log|J_1|.$$
   \qed

   \begin{thm} 	\label{thm:causality_det_map}
	({\bf Property of causality})
   If $\mapping_1$ is independent of $x_2$, then $T_{2\to1}=0$.
   \end{thm}
   \pf By the definition of the F-P operator,
	\begin{eqnarray*}
	\int_{\omega_1} (\FP_\excl2\rho)_1(x_1) dx_1
	&=& \int_{\omega_1\times\R^{n-2}} \FP_\excl2\rho(x_1,x_3,...,x_n)
		dx_1dx_3...dx_n	\\
	&=& \int_{\inmapping_\excl2(\omega_1\times\R^{n-2})}
		\rho_\excl2(x_1,x_3,...,x_n) dx_1dx_3...dx_n	
	\end{eqnarray*}
for any $\omega_1\subset\R$. Note
	$$\inmapping_\excl2(\omega_1\times\R^{n-2})
	  = \inmapping_{1\excl2}(\omega_1 \times \R^{n-2}).$$ 
   That is to say,
	\begin{eqnarray*}
	\int_{\omega_1} (\FP_\excl2\rho)_1(x_1) dx_1
	&=& \int_{\inmapping_{1\excl2}\omega_1} dx_1
	    \int_{\R^{n-2}} \rho_\excl2(x_1,x_3,...,x_n) dx_3...dx_n \\
	&=& \int_{\inmapping_{1\excl2}\omega_1} \rho_1(x_1) dx_1	
	= \int_{\inmapping_1\omega_1} \rho_1(x_1) dx_1
	\end{eqnarray*}
   since $\mapping_1$ (hence $\inmapping_1$) is independent of $x_2$.
   On the other hand,
	\begin{eqnarray*}
	\int_{\omega_1} (\FP\rho)_1(x_1) dx_1
	&=& \int_{\omega_1\times\R^{n-1}} \FP\rho(\ve x) d\ve x	\\
	&=& \int_{\inmapping(\omega_1\times\R^{n-1})} \rho(\ve x) d\ve x \\
	&=& \int_{\inmapping_1\omega_1\times\R^{n-1}} \rho(\ve x) d\ve x \\
	&=& \int_{\inmapping_1\omega_1} dx_1 \int_{\R^{n-1}} 
			\rho(\ve x) dx_2...dx_n		\\
	&=& \int_{\inmapping_1\omega_1} \rho_1(x_1) dx_1.
	\end{eqnarray*}
   So $\int_{\omega_1} (\FP_\excl2\rho)_1(x_1) dx_1 
	= \int_{\omega_1} (\FP\rho)_1 (x_1) dx_1$, $\forall\omega_1\subset\R$,
   and hence $(\FP_\excl2\rho)_1 \overset {a.e.} = (\FP\rho)_1$. Therefore,
	$$ E\log(\FP\rho)_1(x_1) = E\log(\FP_\excl2\rho)_1(x_1),$$
   and
	$$T_{2\to1} = H_1(\tau+1) - H_{1\excl2}(\tau+1) = 0.$$
   \qed

\subsection{Application--Kaplan-Yorke map}
Once a dynamical system is specified, in principle the information flow can
be obtained. This subsection presents an application with a 
discrete-time dynamical system, the Kaplan-Yorke map\cite{Kaplan},
that exhibits chaotic behavior.

\begin{figure} [h]
   \begin{center}
   \includegraphics[angle=0,width=0.5\textwidth] {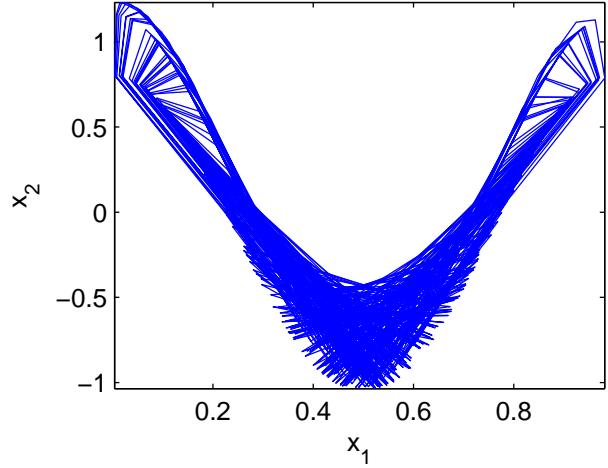}
   \caption
        {The attractor of the Kaplan-Yorke map
	(\ref{eq:kap1})-(\ref{eq:kap2}) with $\alpha=0,2$. 
	To avoid the round-off
	error in the computation which will quickly lead to a zero $x_1$, 
	we let $b=9722377$, and instead compute 
		$a_{n+1} = 2a_n \mod b, \ x_{1,n+1} = a_n / b,\
		 x_{2,n+1} = \alpha x_{2,n} + \cos(4\pi x_{1,n})$.
	The trajectory is initialized with $x_1=7722377/b$, $x_2=0$.
	(The initial points outside the attractor are not shown.)
	}
	\protect{\label{fig:kaplan}}
   \end{center}
   \end{figure}

The Kaplan-Yorke map is defined as a mapping
$\mapping=(\mapping_1,\mapping_2): [0,1]\times\R \to [0,1]\times\R$, 
$(x_1,x_2) \mapsto (y_1, y_2)$, such that
	\begin{eqnarray}
	&& y_1 = \mapping_1(x_1,x_2) = 2x_1 \mod 1, \label{eq:kap1} \\
	&&y_2 = \mapping_2(x_1,x_2) = \alpha x_2 + \cos(4\pi x_1).
						    \label{eq:kap2}
	\end{eqnarray}
A typical trajectory for $\alpha=0.2$ is plotted in Fig.~\ref{fig:kaplan}.
We now compute the information flows between the two components.

First we need to find the F-P operator $\FP\rho(y_1, y_2)$. 
Pick a domain $\omega = [0,y_1] \times [0,y_2]$.
By (\ref{eq:FPcompute})
	\begin{eqnarray}	\label{eq:FPcompute2}
	\FP\rho(y_1,y_2) = \frac{\D^2} {\D y_2 \D y_1} 
	\int_{\inmapping(\omega)}\rho(\xi_1,\xi_2) d\xi_1 d\xi_2,
	\end{eqnarray} 
so the key is the finding of $\inmapping(\omega)$. Since
	\begin{eqnarray*}
	y_1 = \left\{\begin{array}{ll}
			2x_1,   & \qquad 0\le x_1 \le \frac12, \\
			2x_1-1, & \qquad x_1>\frac12
		     \end{array}\right.
	\end{eqnarray*}
it is easy to obtain 
	\begin{eqnarray}
	\mapping_1^{-1}([0,y_1]) = 
	\bracket{0, \frac{y_1}2} \cup \bracket{\frac12, \frac{1+y_1}2}.
	\end{eqnarray}
Given $y_1$, $x_1$ may be either $y_1/2$ or $(1+y_1)/2$, but either way, 
$\cos (4\pi x_1) = \cos(2\pi y_1)$. Thus
	\begin{eqnarray}
	\mapping_2^{-1}(\{y_1\} \times [0,y_2])
	= \bracket{-\frac{\cos 2\pi y_1} \alpha,\ 
		    \frac{y_2 - \cos 2\pi y_1} \alpha}.
	\end{eqnarray}
Eq.~(\ref{eq:FPcompute2}) is, therefore,
	\begin{eqnarray*}
	&& \FP\rho(y_1,y_2) = 
	\frac{\D^2} {\D y_2 \D y_1} \int_0^{y_1/2} d\xi_1 
	\int_{-\frac{\cos2\pi y_1}\alpha}^{\frac{y_2-\cos2\pi y_1} \alpha}
		\rho(\xi_1,\xi_2) d\xi_2	\\
	&&\qquad +
	\frac{\D^2} {\D y_2 \D y_1} \int_{1/2}^{(1+y_1)/2} d\xi_1 
	\int_{-\frac{\cos2\pi y_1}\alpha}^{\frac{y_2-\cos2\pi y_1} \alpha}
		\rho(\xi_1,\xi_2) d\xi_2	\\
	&& =
	\frac 1 {2\alpha} \bracket{
		\rho\parenth{\frac{y_1}2, \frac{y_2-\cos2\pi y_1}\alpha}
	      + \rho\parenth{\frac{1+y_1}2, \frac{y_2-\cos2\pi y_1}\alpha}
				  }		\\
	&& \ \
	+ \frac 1 \alpha \bracket{ \int_0^{y_1/2} \DyI\ 
		\rho\parenth{\xi_1, \frac{y_2-\cos2\pi y_1}\alpha} d\xi_1
	      + 
		\int_{1/2}^{(1+y_1)/2} \DyI\
		\rho\parenth{\xi_1, \frac{y_2-\cos2\pi y_1}\alpha} d\xi_1
				 }.
	\end{eqnarray*} 

To compute $T_{2\to1}$, freeze $x_2$. The resulting mapping
$\mapping_\excl2$ is the dyadic mapping in the $x_1$ direction. As above,
	\begin{eqnarray*}
	\mapping_\excl2^{-1}([0,y_1]) = 
	\bracket{0,\frac{y_1}2} \cup \bracket{\frac12, \frac{1+y_1}2},
	\end{eqnarray*}
which gives
	\begin{eqnarray*}
	\FP_\excl2\rho(y_1) 
	&=& \DyI\ \int_{\mapping_{\excl2}^{-1}([0,y_1])}
		\rho_1(\xi_1) d\xi_1	\\
	&=&
	 \frac12 \bracket{\rho_1\parenth{\frac {y_1} 2} 
	+ \rho_1\parenth{\frac{1+y_1}2}}.
	\end{eqnarray*}
On the other hand, 
	\begin{eqnarray*}
	(\FP)_1(y_1) 
	&=& \int_\R \FP\rho(y_1,y_2) dy_2  \\
	&=& \frac12 \rho_1(\frac{y_1}2) + \frac12 \rho_1(\frac{1+y_1}2)
	+ \int_0^{y_1/2} \DyI\ \rho_1(\xi_1) d\xi_1
	+ \int_{1/2}^{(1+y_1)/2} \DyI\ \rho_1(\xi_1) d\xi_1 \\
	&=&
	\frac12 \bracket{\rho_1\parenth{\frac{y_1} 2}
			+ \rho_1\parenth{\frac{1+y_1} 2 }}.
	\end{eqnarray*}
So
	\begin{eqnarray}
	T_{2\to1} = E\log(\FP_\excl2\rho)_1(y_1)
		  - E\log(\FP\rho)_1(y_1) = 0,
	\end{eqnarray}
just as one would expect based on the independence of $\mapping_1$ on
$x_2$. This serves as a validation of Theorem~\ref{thm:causality_det_map}.

To compute $T_{1\to2}$, notice
	\begin{eqnarray*}
	\mapping_{\excl1}^{-1}([0,y_2]) = 
	\bracket{-\frac{\cos4\pi x_1} \alpha,\  
		  \frac{y_2 - \cos4\pi x_1} \alpha}.
	\end{eqnarray*}
The corresponding F-P operator is such that
	\begin{eqnarray*}
	(\FP_\excl1\rho)(y_2) = 
\DyII\ \int_{-\frac{\cos4\pi x_1} \alpha}^{\frac{y_2 - \cos4\pi x_1} \alpha}
	\rho_2(\xi_2) d\xi_2 	
	= \frac 1 \alpha \rho_2\parenth{\frac{y_2 - \cos4\pi x_1} \alpha}
	= \frac1\alpha \rho_2(x_2),
	\end{eqnarray*}
which makes sense, considering that, when $x_1$ is frozen,
$y_2$ is just a translation followed by a rescaling of $x_2$.
On the other hand, the marginal density
	\begin{eqnarray*}
	&&(\FP\rho)_2(y_2) = \int_0^1 \FP\rho(y_1, y_2) dy_1 \\
	&& = 
	   \frac1{2\alpha} \int_0^1 \bracket{
		\rho\parenth{\frac{y_1}2, \frac{y_2-\cos2\pi y_1}\alpha}
	  + \rho\parenth{\frac{1+y_1}2, \frac{y_2-\cos2\pi y_1}\alpha}
					   }\ dy_1	\\
	&&\ \ 
	  + \frac1\alpha \int_0^1 dy_1 \bracket{
	  \int_0^{y_1/2} \DyI\ \rho\parenth{\xi_1, 
		\frac{y_2-\cos2\pi y_1}\alpha} d \xi_1
	  +
 	 \int_{1/2}^{(1+y_1)/2} \DyI\ \rho\parenth{\xi_1, 
		\frac{y_2-\cos2\pi y_1}\alpha} d \xi_1
					   }.
	\end{eqnarray*}
Because of the intertwined $y_1$ and $y_2$, these integrals 
cannot be explicitly evaluated without specifications of $\rho$. 
But when $\rho$ is given, it is a straightforward exercise to compute
	$$-E\log(\FP\rho)_2(y_2) = 
	-\int_0^1\int_\R \log(\FP\rho)_2(\mapping_2(x_1,x_2))
	\rho(x_1, x_2) dx_1 dx_2.$$
Denote it by $\tilde H_2$. Then
	\begin{eqnarray}
	T_{1\to2} 
	&=& E\log(\FP_\excl1\rho)_2(\mapping_2(x_1,x_2))
		  - E\log(\FP\rho)_2(\mapping_2(x_1,x_2))	\cr
	&=& \int_0^1\int_\R \frac1\alpha \rho_2(x_2) \rho(x_1,x_2) dx_1 dx_2
		  + \tilde H_2					\cr
	&=&
	    \tilde H_2 - H_2/\alpha.
	\end{eqnarray}
Generally this does not vanish. That is to say, within the Kaplan-Yorke
map, there exists a one-way information flow from $x_1$ to $x_2$.

\subsection{Applications--The baker transformation and \Henon\ map revisited}

Since its establishment, the Liang-Kleeman formalism has been
applied to a variety of dynamical system problems. Hereafter we will
re-study some benchmark examples and see whether the results are different.
In this subsection we look at the baker transformation and H\'enon\ map.


\subsubsection{Baker transformation}

\begin{figure}[h]
	\begin{center}
	\includegraphics[angle=0, width=1\textwidth]{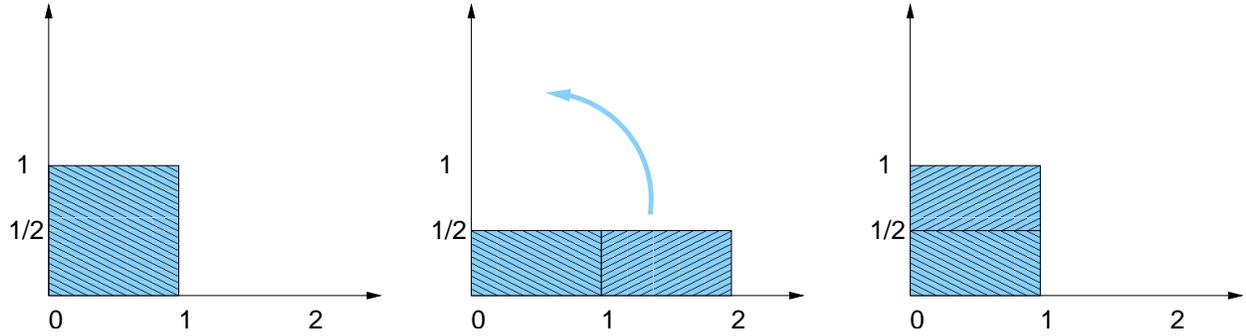}
	\caption{A schematic of the unidirectional information
		flow from the abscissa to the ordinate 
		upon applying the baker transformation.
		\protect{\label{fig:baker}}}
	\end{center}
	\end{figure}

The baker transformation is an extensively studied 
prototype of area-conserving chaotic maps that has been 
used to model the diffusion process in real physical world. It
mimicks the kneading of dough:
first the dough is compressed, then cut in half; the two halves
are stacked on one-another, compressed, and so forth; see
Fig.~\ref{fig:baker} for an illustration. In formal language,
it is $\mapping:\Omega\to \Omega$, $\Omega=[0,1]\times[0,1]$ being a unit
square,
        \begin{eqnarray}	\label{eq:bakerdefn}
        \mapping(x_1,x_2) = \left\{\begin{array}   {ll}
               (2x_1,\ \frac {x_2} 2),\qquad   
			& 0\le x_1 \le \frac 1 2, \ 0\le x_2 \le 1,  \\
               (2x_1-1,\ \frac 1 2 x_2 + \frac 1 2), \qquad\qquad
                        & \frac 1 2<x_1\le 1, \ 0\le x_2 \le 1.
                        \end{array}\right.
        \end{eqnarray}
It is invertible, and the inverse is
        \begin{eqnarray}
        \mapping^{-1}(x_1,x_2) = \left\{\begin{array}{l l}
                (\frac {x_1} 2,\ 2x_2),\qquad &        0\le x_2\le \frac 1 2,\
                                                0\le x_1 \le 1,    \\
                (\frac{x_1+1} 2,\ 2x_2-1),\qquad\qquad   
					& \frac 1 2 \le x_2 \le 1,\
                                                0\le x_1 \le 1.
                      \end{array}\right.
        \end{eqnarray}
Thus the F-P operator $\FP$ can be easily found
        \begin{eqnarray}	\label{eq:FP_baker}
        \FP\rho(x_1,x_2) 
		= \rho\bracket{\mapping^{-1}(x_1,x_2)}\cdot\abs{J^{-1}}
		= \left\{\begin{array}{ll}
                \rho(\frac {x_1} 2, 2x_2),\qquad       
					&0\le x_2<\frac 1 2,  \\
                \rho(\frac {1+x_1}2, 2x_2-1), \qquad
					&\frac 1 2\le x_2\le 1.
                \end{array}\right.
        \end{eqnarray}

We now use the above theorem to compute $T_{2\to1}$.
Integrating (\ref{eq:FP_baker}) with respect to $x_2$, 
	\begin{eqnarray}	\label{eq:marginal}
	(\FP\rho)_1(x_1)
	&=& \int_0^{1/2} \rho(\frac {x_1} 2, 2x_2)\ dx_2 +
	    \int_{1/2}^1 \rho(\frac {x_1+1} 2, 2x_2-1)\ dx_2	\cr
	&=& \frac 1 2 \int_0^1 \bracket{
		\rho\parenth{\frac {x_1} 2,x_2} + 
		\rho\parenth{\frac {x_1+1} 2, x_2}
		}\ dx_2						\cr
	&=& \frac 1 2 \bracket{
		\rho_1\parenth{\frac {x_1} 2} +
		\rho_1\parenth{\frac {x_1+1} 2}
		}.
	\end{eqnarray}
When $x_2$ is frozen as a parameter, the baker transformation
(\ref{eq:bakerdefn}) becomes a dyadic mapping in $x_1$ direction,  i.e.,
a mapping $\mapping_1: [0,1] \to [0,1]$, 
	\begin{eqnarray*}
	\mapping_1(x_1) = 2x_1\ (\rm mod\ 1).
	\end{eqnarray*}
For any $0<x_1<1$, The counterimage of $[0,x_1]$ is
	\begin{eqnarray*}
	\mapping^{-1}([0,x_1]) = \bracket{0, \frac {x_1} 2} \cup 
				 \bracket{\frac 1 2, \frac {1+x_1} 2}.
	\end{eqnarray*}
So
	\begin{eqnarray*}
	(\FP_{\excl2}\rho)_1(x_1) 
	&=& \frac {\D} {\D x_1} \int_{\mapping^{-1}([0,x_1])} \rho(s)\ ds \\
	&=& \frac {\D} {\D x_1} \int_0^{x_1/2} \rho(s)\ ds +	
	    \frac {\D} {\D x_1} \int_{1/2}^{(1+x_1)/2} \rho(s)\ ds	  \\
	&=& \frac 1 2 \bracket{
		\rho\parenth{\frac {x_1} 2}  + \rho\parenth{\frac {1+x_1} 2}}.
	\end{eqnarray*}
Thus
	\begin{eqnarray*}
	(\FP_\excl2\rho)_1(x_1) = (\FP\rho)_1(x_1).
	\end{eqnarray*}
By the above theorem,
	\begin{eqnarray*}
	T_{2\to1} = E\log(\FP_\excl2\rho)_1(\mapping_1(\ve x))
		  - E\log(\FP\rho_1(\mapping_1(\ve x)) = 0.
	\end{eqnarray*}

To compute $T_{1\to2}$, observe
	\begin{eqnarray}
	(\FP\rho)_2(x_2) = 
  \int_0^1 \FP\rho(x_1,x_2)\ dx_1 
	= \left\{\begin{array}{ll}
	\int_0^1 \rho\parenth{\frac {x_1} 2, 2x_2}\ dx_1,\qquad	
		&	0 \le x_2 < \frac 1 2;			\\
	\int_0^1 \rho\parenth{\frac {x_1+1} 2, 2x_2-1}\ dx_1, \qquad
		& \frac 1 2 \le x_2 \le 1.
				    \end{array}\right.
	\end{eqnarray}
By Proposition~\ref{thm:prop3.1},
	\begin{eqnarray*}
	&& H_2(\tau+1) = -E_x \log(\FP\rho)_2(\mapping_2(x_1,x_2))\\
	&&= - \int_0^{1/2} \rho_2(x_2) 
	      \log\parenth{\int_0^1\rho(\frac\lambda 2, x_2) d\lambda} dx_2
	    - \int_{1/2}^1 \rho_2(x_2) 
	      \log\parenth{\int_0^1\rho(\frac{\lambda+1}2,x_2)d\lambda} dx_2\\
	&&= - \int_0^{1/2} \rho_2(x_2) 
	      \log\parenth{2\int_0^{1/2} \rho(\xi, x_2) d\xi} dx_2
	    - \int_{1/2}^1 \rho_2(x_2) 
	      \log\parenth{2\int_{1/2}^1 \rho(\xi, x_2) d\xi} dx_2 \\
	&&= -\log2 -
	      \int_0^{1/2} \rho_2(x_2) 
	      \log\parenth{\int_0^{1/2} \rho(\xi, x_2) d\xi} dx_2
	    - \int_{1/2}^1 \rho_2(x_2) 
	      \log\parenth{\int_{1/2}^1 \rho(\xi, x_2) d\xi} dx_2, 
	\end{eqnarray*}
so 
  \begin{eqnarray*}
  \Delta H_2 &=&  H_2(\tau+1) - H_2(\tau) \\
	&=& -\log2 -
	      \int_0^{1/2} \rho_2(x_2) 
	      \log\parenth{\int_0^{1/2} \rho(\xi, x_2) d\xi} dx_2
	    - \int_{1/2}^1 \rho_2(x_2) 
	      \log\parenth{\int_{1/2}^1 \rho(\xi, x_2) d\xi} dx_2 \\
	&\ & + \int_0^1\int_0^1 \rho(x_1,x_2)\cdot
     \bracket{\log\parenth{\int_0^1\rho(\lambda,x_2)d\lambda }} dx_1dx_2\\
	&=& - \log 2 + (I + \II),
	\end{eqnarray*}
where
	\begin{eqnarray}
	 I &=& \int_0^{1/2} \rho_2(x_2)\cdot\bracket{
		\log \frac {\int_0^1\rho(\lambda,x_2)d\lambda}	
		   {\int_0^{1/2}\rho(\lambda,x_2)d\lambda} }\ dx_2, \\
	\II &=& \int_{1/2}^1 \rho(x_2)\cdot\bracket{
		\log \frac {\int_0^1\rho(\lambda,x_2)d\lambda}	
		   {\int_{1/2}^1\rho(\lambda,x_2)d\lambda} }\ dx_2. 
	\end{eqnarray}

To compute $H_{2\excl1}$, notice that, when $x_1$ is frozen,
the transformation is invertible; moreover,
the Jacobian $J_2 = 1/2$ is a constant. By Theorem~\ref{thm:2d_dis},
	\begin{eqnarray}
	\Delta H_{2\excl1} = E\log\frac 1 2 = - \log 2,
	\end{eqnarray}
which gives
	\begin{eqnarray}	\label{eq:baker_T12}
	T_{1\rightarrow 2} = \Delta H_2 - \Delta H_{2\excl1}
			   = I + \II.
	\end{eqnarray}
It is easy to show that $I + \II > 0$. In fact, obviously $I+\II$ is
nonnegative; besides, the two brackets cannot be zero simultaneously, so
it cannot be zero. Hence $T_{1\to2}$ is strictly 
positive; that is to say, 
there is always information flowing from the abscissa to the ordinate.

To summarize, $T_{2\to1} = 0$, $T_{1\to2} = I + \II > 0$.
These results are precisely the same as those obtained before in
\cite{LK05} and \cite{LK07a}. 
That is to say, for the baker transformation, the current formalism shows no
difference from the previous one based on heuristic arguments and with
approximations.


\subsubsection{\Henon\ map}

\begin{figure}[h]
	\begin{center}
	\includegraphics[angle=0, width=0.5\textwidth]{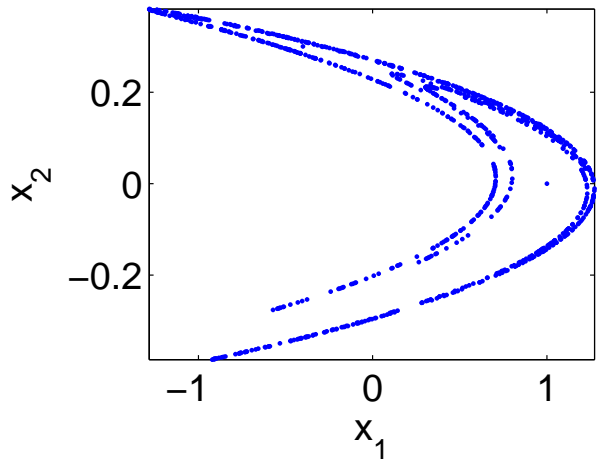}
	\caption{A typical trajectory of the canonical 
		\Henon\ map ($a=1.4$, $b=0.3$).
		\protect{\label{fig:henon}}}
	\end{center}
	\end{figure}

The \Henon\ map is a mapping
	$\mapping=(\mapping_1,\mapping_2): \R^2 \mapsto \R^2$
defined such that
	\begin{eqnarray}	\label{eq:henon}
	\left\{
	\begin{array}{l}
	\mapping_1(x_1,x_2) = 1 + x_2 - a x_1^2,	\\
	\mapping_2(x_1,x_2) = bx_1,
	\end{array}
	\right.
	\end{eqnarray}
with $a>0$, $b>0$.  The case with parameters
$a=1.4$ and $b=0.3$ is called a ``canonical \Henon\ map,'' whose
attractor is shown in Fig.~\ref{fig:henon}.

It is easy to see that the \Henon\ map is invertible; its inverse
is
	\begin{eqnarray}
	\mapping^{-1}(x_1, x_2) = 
		\parenth{\frac {x_2} b,\ x_1 - 1 + \frac a {b^2} x_2^2}.
	\end{eqnarray}
The F-P operator thus can be easily found from (\ref{eq:FP_explicit}):
	\begin{eqnarray}	\label{eq:FP_henon}
	\FP\rho(x_1,x_2) &=& \rho(\mapping^{-1}(x_1,x_2)) |J^{-1}|\cr
      &=& \frac 1 b\cdot 
	  \rho\parenth{\frac {x_2} b,\ x_1 - 1 + \frac a {b^2} x_2^2}.
	\end{eqnarray}
In the following we compute the flows between the quadratic component
$x_1$ and the linear component $x_2$. 

Look at $T_{2\to1}$ first.
By (\ref{eq:T21_det_dis}),
we need to find the marginal density of $x_1$ at step $\tau+1$ with and
without the effect of $x_2$, i.e., $(\FP\rho)_1$ and 
$(\FP\rho)_{1\excl2}$. From (\ref{eq:FP_henon}), 
$(\FP\rho)_1$ is
	\begin{eqnarray*}
	(\FP\rho)_1(x_1)  
	&=& \int_\R \FP\rho (x_1, x_2)\ dx_2		\\
	&=& \int_\R \frac 1 b  \cdot \rho
		\parenth{\frac{x_2}b, x_1-1+\frac ab x_2^2}\ dx_2 \\
	&=& \int_\R \rho(\eta, x_1-1+ a \eta^2)\ d\eta.
		\qquad\qquad (x_2/b \equiv \eta)
	\end{eqnarray*}
If $a=0$, this would give $\rho_2(x_1-1)$, i.e., the
marginal pdf of $x_2$ with argument $x_1-1$. 
But here $a>0$, the integration is taken along a parabolic curve rather
than a straight line. Still the final result will be related to the
marginal density of $x_2$; for notational simplicity, write
	\begin{eqnarray}
	(\FP\rho)_1(x_1) = \tilde\rho_2(x_1).
	\end{eqnarray}

To find $(\FP_{\excl2}\rho)_1$, use $y_1$ to denote
	\begin{eqnarray*}
	\mapping_1(x_1) = 1 + x_2 - a x_1^2,
	\end{eqnarray*}
following our convention to distinguish variables at different steps.
Modify the system so that $x_2$ is now a parameter. As before, we need to
find the counterimage of $(-\infty, y_1]$ under the transformation
with $x_2$ frozen:
	\begin{eqnarray*}
	 \mapping_1^{-1} ((-\infty,y_1])
	 = \left({-\infty,\ \ -\sqrt{{(1+x_2-y_1)} / a}}\right] 
	   \cup 
	   \left[{\sqrt{{(1+x_2-y_1)}/a},\ \ \infty}\right).
	\end{eqnarray*}
Therefore,
	\begin{eqnarray*}
	&&(\FP_{\excl2}\rho)_1 (y_1) = \frac {d} {dy_1} 
	    \int_{\mapping_1^{-1} ((-\infty,y_1])} \rho_1(s)\ ds \\
	&&\ \ = \frac {d} {dy_1} 
	    \int_{-\infty}^{-\sqrt{{(1+x_2-y_1)}/a}} 
		\rho_1(s)\ ds +
 	    \frac {d} {dy_1} 
	    \int_{\sqrt{{(1+x_2-y_1)}/a}}^{\infty} 
		\rho_1(s)\ ds 				    \\
	&&\ \ = \frac 1 {2\sqrt{a(1+x_2-y_1)}} \bracket{
		\rho_1\parenth{-\sqrt{{(1+x_2-y_1)}/ a}} +
		\rho_1\parenth{\sqrt{(1+x_2-y_1)/a}} }		\\
	&& 	\qquad\qquad\qquad\qquad\qquad\qquad\qquad\qquad\qquad
			\qquad\quad (y_1<1+x_2)	\\
	&&\ \ = \frac 1 {2a |x_1|} \bracket{\rho_1(-x_1) + \rho_1(x_1)}.
			\qquad\quad ({\rm recall}\ y_1 = 1+x_2-ax_1^2)
	\end{eqnarray*}
Denote the average of $\rho_1(-x_1)$ and $\rho_1(x_1)$ as $\bar\rho_1(x_1)$
to make an even function of $x_1$.  Then $\FP_{\excl2}\rho)_1$ is simply 
	\begin{eqnarray}
	(\FP_{\excl2}\rho)_1(y_1) = \frac {\bar\rho_1(x_1)} {a|x_1|}.
	\end{eqnarray}
Note that the parameter $x_2$ does not appear in the arguments. 
Substitute all the above into (\ref{eq:T21_det_dis}) to get
	\begin{eqnarray*}
	T_{2\to1} 
	&=&  E\log(\FP_\excl2\rho)_1(y_1) - E\log(\FP\rho)_1(y_1) \\
	&=& E\log \frac{\bar\rho_1(x_1)} {a|x_1|}
	    - E\log\tilde\rho_2(1+x_2-ax_1^2) \\
	&=& E\log\bar\rho_1(x_1) - E\log|ax_1| 
	    - E\log\tilde\rho_2(1+x_2-ax_1^2).
	\end{eqnarray*}
Comparing this to the result in \cite{LK07a}, except for the term
$-E\log|ax_1|$, all other terms are different.

Next consider $T_{1\to2}$.
From (\ref{eq:FP_henon}), the marginal density of $x_2$ at $\tau+1$ is
	\begin{eqnarray*}
	(\FP\rho)_2(x_2) 
	&=& \int_\R \FP\rho(x_1, x_2)\ dx_1	\\
	&=& \int_\R \frac 1 b 
	    \rho\parenth{\frac {x_2} b,\ 
			 x_1 - 1 + a \frac {x_2^2} {b^2}}\ dx_1 \\
	&=& \frac 1 b \int_\R \rho(y, \xi)\ d\xi
	= \frac 1 b \rho_1\parenth{\frac {x_2} b}.
	\end{eqnarray*}
Thus
	\begin{eqnarray*}
	H_2 &=& -E(\FP\rho)_2(y_2) \cr
	&=& - \int_\R \frac 1 b \rho_1\parenth{\frac {x_2} b} \cdot
	      \log\bracket{\frac 1 b \rho_1\parenth{\frac {x_2} b}}\ dx_2  \cr
	&=& H_1 + \log b.
	\end{eqnarray*}
The evaluation of $H_{2\excl1}$ is much easier. 
As $x_1$ is frozen as a parameter, $y_2$ becomes definite.
In this case, the 2D random variable is degenerated to a 1D variable.
Correspondingly $\FP_\excl1\rho$ becomes a pdf in $x_1$ only. So
	$$(\FP_\excl1\rho)_2 = \int_\R \FP_\excl1\rho dx_1 = 1.$$
Thus $H_{2\excl1} = -E\log(\FP_\excl1\rho)_2 = 0$.
By (\ref{eq:T21_det_dis}), the information flow from $x_1$ to $x_2$
is, therefore, 
	\begin{eqnarray}	\label{eq:henon_T12}
	T_{1\to2} = H_2 - H_{2\excl1} = H_1 + \log b.
	\end{eqnarray}
In other words, the flow from $x_1$ to $x_2$ is equal to the marginal 
entropy of $x_1$, modified by an amount related to the factor $b$.
Particularly, when $b=1$, $T_{1\to2} = H_1$.
This is precisely the same as what is obtained before in \cite{LK07a}.

%

In a summary, the information flows within the baker transformation 
are precisely the same as we have obtained before in \cite{LK07a}.
For the \Henon\ map, the flow from $x_1$ to $x_2$, has recovered 
the benchmark result based on physical grounds. But 
$T_{2\to1}$ is generally different from that in \cite{LK07a}.

\section{Continuous-time deterministic systems}	\label{sect:det_flow}

\subsection{Deriving the information flow}

Now look at the information flow within the continuous system, the 2D
version of which motivates this line of work:
	\begin{eqnarray}
	&& \dt {x_1} = F_1(t; x_1,x_2,...,x_n),   \label{eq:gov1} \\
	&& \dt {x_2} = F_2(t; x_1,x_2,...,x_n),   \label{eq:gov2} \\
	&& \quad \vdots \qquad\quad\qquad \vdots	     \\
	&&\dt {x_n} = F_n(t; x_1,x_2,...,x_n),    \label{eq:gov3}
	\end{eqnarray}
or, in vectorial form,
	\begin{eqnarray}
	\dt {\ve x} = \ve F(t; \ve x).
	\end{eqnarray}
Consider a time interval $[t, t+\delt]$. Following \cite{LK07b}, we
discretize the ordinary differential equation and construct a mapping
	$\mapping: \R^n \to \R^n$ , 
	$\ve x(t) \mapsto \ve x(t+\delt) = \ve x + \ve F\delt$.
Correspondingly there is a Frobenius-Perron operator
	$\FP: L^1(\R^n) \to L^1(\R^n)$,
	$\rho(t) \mapsto \rho(t+\delt)$.
Write $\ve x(t+\delt)$ as $\ve y$, a convention we have been using all
the time to avoid confusion. Then the mapping $\mapping: \ve x \mapsto \ve
y$ is such that
	\begin{eqnarray}
	\left\{\begin{array}{l}
	 y_1 = x_1 + F_1(x_1,x_2,...,x_n) \delt,		\\
	 y_2 = x_2 + F_2(x_1,x_2,...,x_n) \delt,		\\
	 \quad\ \vdots \qquad\qquad \vdots \cr
	 y_n = x_n + F_n(x_1,x_2,...,x_n) \delt.
	\end{array}\right.
	\end{eqnarray}
Its Jacobian is
	\begin{eqnarray}
	J &=& \det\parenth{\frac{\D\ve y}{\D\ve x}}
	  = \det\matthree {1+\DI{F_1}\delt} {\hdots} {\Dn{F_1}\delt}
			  {\vdots} {\ddots} {\vdots}
			  {\Dn{F_n}\delt} {\hdots} {1+\Dn{F_n}\delt} \cr
	&=& 1 + \sum_i \Di {F_i} \delt + \hot	\cr
	&=& 1 + \nabla\cdot\ve F \delt + \hot.
	\end{eqnarray}
As $\delt\to0$, $J\to1\ne0$, so $\mapping$ thus constructed is always
invertible for $\delt$ small enough. Moreover, it is easy to obtain 
the inverse mapping 
	\begin{eqnarray}
	\inmapping: \ve x = \ve y - \ve F\delt + \hot
	\end{eqnarray}
and $J^{-1} = 1 - \nabla\cdot\ve F\delt + \hot$.
So 
	\begin{eqnarray*}
	\FP\rho(\ve y) 
	&=& \rho(\inmapping(\ve y)) \cdot |J^{-1}| \\
	&=& \rho(\ve y - \ve F\delt) \cdot (1 - \nabla\cdot\ve F\delt)+\hot\\
	&=& \rho(\ve y) - \nabla\rho\cdot\ve F\delt 
		   	- \rho\nabla\cdot\ve F\delt + \hot \\
	&=& \rho(\ve y) - \nabla\cdot(\rho\ve F) \delt + \hot.
	\end{eqnarray*}
As a verification, check
	\begin{eqnarray*}
	\Dt\rho = \lim_{\delt\to0} \frac {\FP\rho(\ve x) - \rho(\ve x)} \delt
	= -\nabla \cdot (\rho\ve F).
	\end{eqnarray*}
This yields the Liouville equation $\Dt\rho + \nabla\cdot(\rho\ve F) = 0$, 
as is expected.

With $\FP\rho$ we now can compute the marginal density 
	\begin{eqnarray*}
	(\FP)_1(y_1) = \rho_1(y_1) - \delt\int_{R^{n-1}} \DyI{\rho F_1}
		dy_2...dy_n + \hot
	\end{eqnarray*}
which gives
	\begin{eqnarray*}
	-\log(\FP\rho)_1(y_1) 
	&=& -\log\rho_1(y_1) -
	  \bracket{\log \parenth{1 - \frac\delt{\rho_1} 
	   	\int_{\R^{n-1}} \DyI{\rho F_1} dy_2...dy_n + \hot}}	\\
	&=& - \log\rho_1(y_1)  + \frac\delt{\rho_1(y_1)} 
	   	\int_{\R^{n-1}} \DyI{\rho F_1} dy_2...dy_n + \hot	\\
	&=& - \log\rho_1(x_1+F_1\delt)  + \frac\delt{\rho_1(x_1)} 
	   	\int_{\R^{n-1}} \DI{\rho F_1} dx_2...dx_n + \hot.
	\end{eqnarray*}
At the last step, the $y's$ have been replaced by $x's$ in the integral
term. This is legitimate since the difference goes to higher order terms.

We now evaluate the marginal entropy increase at $t+\delt$.
The following is the key step: Take expectation on both sides, the left
hand side with respect to $(\FP\rho)_1(y_1)$, while the right hand side
with respect to $\rho_1(x_1)$. This yields
	\begin{eqnarray*}
	H_1(t+\delt) 
	&=& -E \log\rho_1(x_1+F_1\delt) + 
	    \delt E \parenth{\frac 1{\rho_1} \int_{\R^{n-1}} 
		\DI{\rho F_1} dx_2...dx_n} + \hot.	\cr
	&=& H_1(t) - E \DI {\rho\rho_1} F_1\delt
	    + \delt \int_\R \rho_1 \frac 1{\rho_1} dx_1
		    \int_{\R^{n-1}} \DI {\rho F_1} dx_2...dx_n + \hot.
	\end{eqnarray*}
Note the third term on the right hand side vanishes after integration with
respect to $x_1$ due to the compactness of the functions. So
	\begin{eqnarray*}
	H_1(t+\delt) = H_1(t) - \delt E\parenth{F_1\DI {\log\rho_1}} + \hot,
	\end{eqnarray*}
and hence
	\begin{eqnarray}
	\dt {H_1} = \lim_{\delt\to0} \frac {H_1(t+\delt) - H_1(t)} \delt
	 = - E\parenth{F_1\DI {\log\rho_1}}.
	\end{eqnarray}
This is precisely the same as that either from the F-P operator\cite{LK07a} 
or directly from the Liouville equation\cite{LK05}, serving a validation of
our approach in this study.

\vskip 1cm

When $x_2$ is frozen as a parameter on $[t,t+\delt]$, we need to examine
the modified mapping 
	$\mapping_{\excl2}: \R^{n-1} \to \R^{n-1}$
	\begin{eqnarray}
	\left\{\begin{array}{l}
	 y_1 = x_1 + F_1(x_1,x_2,...,x_n) \delt,		\\
	 y_3 = x_3 + F_3(x_1,x_2,...,x_n) \delt,		\\
	 \quad\ \vdots \qquad\qquad \vdots \cr
	 y_n = x_n + F_n(x_1,x_2,...,x_n) \delt,
	\end{array}\right.
	\end{eqnarray}
i.e., the mapping $\mapping$ with the equation 
	$y_2 = x_2 + F_2\delt$
removed, and $x_2$ frozen as a parameter.
Again, here $x_i$ stands for $x_i(t)$, and $y_i$ for $x_i(t+\delt)$.
For convenience, we further adopt the following notations:
	\begin{eqnarray*}
	&& \ve y_{\excl2} = (y_1,y_3,...,y_n)^T, \\
	&& \ve x_{\excl2} = (x_1,x_3,...,x_n)^T, \\
	&& \ve F_{\excl2} = (F_1,F_3,...,F_n)^T. 
	\end{eqnarray*}
Besides, use $\rho_{\excl2}$ to signify the joint density of $\ve
x_{\excl2}$, and $\rho_{1\excl2}$ to denote the density of $x_1$ with
$x_2$ frozen as a parameter on $[t,t+\delt]$. 
Notice the fact $\rho_{1\excl2} = \rho_1$ at time $t$.

It is easy to know that the Jacobian of $\mapping_\excl2$ 
	\begin{eqnarray}
	J_\excl2 = \det \parenth{\frac{\ve y_\excl2} {\ve x_\excl2}}
		 = 1 + \delt \sum_{i\ne2} \Di {F_i} + \hot.
	\end{eqnarray}
The corresponding F-P operator 
	$\FP_\excl2: L^1(\R^{n-1} \to L^1(\R^{n-1}$
is such that
	\begin{eqnarray*}
	\FP_\excl2\rho_\excl2(\ve y) 
	&=& \rho_\excl2(\inmapping_\excl2(\ve y)) \cdot |J_\excl2^{-1}| \\
	&=& \rho_\excl2(\ve y_\excl2 - \ve F_\excl2 \delt) \cdot
		\parenth{1 - \sum_{i\ne2} \Di{F_i} \delt} + \hot 	\\
	&=& \rho_\excl2(\ve y_\excl2) - \nabla\cdot (\rho_\excl2 
				\ve F_\excl2) \delt + \hot.
	\end{eqnarray*}
Integrate with respect to $(y_3,...,y_n)$ (recall that $x_2$ is now a
parameter) to get
	\begin{eqnarray}
	(\FP_\excl2\rho_\excl2)_1(y_1) = \rho_{1\excl2}(y_1)
		- \delt \int_{\R^{n-2}} \DyI {\rho_\excl2 F_1} dy_3...dy_n
		+ \hot,
	\end{eqnarray}
where other terms vanish due to the compactness assumed for the functions.
Hence
	\begin{eqnarray*}
	-\log(\FP_\excl2\rho_\excl2)_1(y_1) 
	&=& - \log\rho_{1\excl2}(y_1) 
	    - \log\parenth{1 - \frac\delt {\rho_{1\excl2}(y_1)}
			  \int \DyI {\rho_\excl2 F_1} dy_3...dy_n}
	    + \hot	\\
	&=& - \log\rho_{1\excl2}(x_1 + F_1\delt) 
	    + \frac\delt {\rho_{1\excl2}(x_1)}
			  \int \DI {\rho_\excl2 F_1} dx_3...dx_n
	    + \hot.
	\end{eqnarray*}
Note in the integral term, the $y's$ have been replaced by $x's$; this is
legitimate as the difference goes to the higher order terms.
Since $\rho_{1\excl2}(x_1) = \rho_1(x_1)$ at $t$, so
	\begin{eqnarray*}
	\rho_{1\excl2}(x_1+F_1\delt) = \rho_1(x_1) + \DI{\rho_1} F_1\delt + \hot
	\end{eqnarray*}
and hence
	\begin{eqnarray*}
	-\log(\FP_\excl2\rho_\excl2)_1(y_1) 
	= - \log\rho_1 - 
		\frac 1 {\rho_1} \DI{\rho_1} F_1 \delt 
	    + \frac\delt {\rho_1(x_1)}
			  \int \DI {\rho_\excl2 F_1} dx_3...dx_n + \hot.
	\end{eqnarray*}
Take expectation on both sides, the left hand side with respect to the
joint probability density of $(y_1,x_2)$, while the right hand side with
respect to $(x_1,x_2)$. This is the key step that makes the present study
fundamentally different from \cite{LK07b} which relies on an approximation
to fulfill the derivation. This yields
	\begin{eqnarray*}
	H_{1\excl2}(t+\delt) 
	&=& H_t(t) -\delt E\parenth{F_1\DI {\log\rho_1}}
	+ \delt \int_{\R^2} \frac{\rho_{12}(x_1,x_2)}{\rho_1(x_1)} dx_1dx_2
		\int_{\R^{n-2}} \DI {\rho_\excl2 F_1} dx_3...dx_n + \hot\\
	&=& H_t(t) -\delt E\parenth{F_1\DI {\log\rho_1}}
	+ \delt \int_\R \rho_{2|1} \DI {\rho_\excl2 F_1} d\ve x + \hot,
	\end{eqnarray*}
where $\rho_{2|1}$ is the conditional density of $x_2$ on $x_1$. Thus
	\begin{eqnarray}
	\dt {H_{1\excl2}} = - E\parenth{F_1\DI {\log\rho_1}}
		    + \int_{\R^n} \rho_{2|1} \DI {\rho_\excl2 F_1} d\ve x.
	\end{eqnarray}
We therefore arrive at the following theorem:
   	\begin{thm}		\label{thm:T21_det_cont}
	\begin{eqnarray}	\label{eq:T21_det_cont}
	T_{2\to1} = \dt {H_1} - \dt {H_{1\excl2}}
	= -\int_{\R^n} \rho_{2|1} \DI {\rho_\excl2 F_1} d\ve x
	= - E\bracket{\frac 1 {\rho_1} 
		\int_{\R^{n-2}} \DI {F_1\rho_\excl2} dx_3...dx_n}.
	\end{eqnarray}
	\end{thm}

\subsection{An alternative derivation}

For the continuous system
	\begin{eqnarray}
	\dt {\ve x} = \ve F(\ve x, t),
	\end{eqnarray}
consider an interval $[t, t+\delt]$, and a mapping
	$$\mapping: \R^n \to \R^n, \qquad
	 \ve x(t) \mapsto \ve x(t+\delt) = \ve x(t) + \ve F\delt.$$
Recall that by definition 
	$E\psi(\ve x(t+\delt) = \int \psi(\ve x)\rho(x,t+\delt) d\ve x,
	+\hot,$
for any test function $\psi$, and 
	\begin{eqnarray*}
	E\psi(\ve x(t+\delt) 
	&=& E\psi(\ve x(t) + \ve F\delt + \hot)\\
	&=& E\bracket{\psi(\ve x(t)) + \nabla\psi\cdot\ve F\delt + \hot}.
	\end{eqnarray*}
Note the expectation on the left hand side is with respect to
$\rho(t+\delt)$, and that on the right is with respect to $\rho(t)$. This
way we obtain the Liouville equation.

Now let $\psi$ be the functional $\log (\FP_\excl2\rho)_1$. When $x_2$ is
frozen, on interval $[t, t+\delt]$, there is a Liouville equation
	\begin{eqnarray}
	\Dt {\rho_\excl2} + \DI {F_1\rho_\excl2} 
	+ \DIII {F_3 \rho_\excl2} + ... + \Dn {F_n \rho_\excl2} = 0
	\end{eqnarray}
for  $\rho_\excl2$ the joint density of $(x_1, x_3,...,x_n)$.
The equation for its marginal density 
	$\rho_{1\excl2} = \int_{\R^{n-2}} \rho_\excl2 dx_3...dx_n$
is, after integration with respect to $(x_3, x_4,...,x_n)$ and with the
consideration of the compact support assumption,
	\begin{eqnarray*}
	\Dt {\rho_{1\excl2}} + \DI\ \int_{\R^{n-2}} F_1\rho_\excl2
			dx_3...dx_n = 0.
	\end{eqnarray*}
Divided by $\rho_{1\excl2}$, this yields
	\begin{eqnarray*}
	\Dt {\log\rho_{1\excl2}} + \int_{\R^{n-2}} 
		\frac 1 {\rho_{1\excl2}} \DI{F_1\rho_\excl2} dx_3...dx_n
	= 0.
	\end{eqnarray*}
Discretizing, and noticing the fact $\rho_{1\excl2}(t) = \rho_1(t)$,
	\begin{eqnarray*}
	\log\rho_{1\excl2}(t+\delt; x_1) = \log\rho_1(t; x_1) 
		- \delt \int_{R^{n-2}} \frac 1 {\rho_{1\excl2}}
		\DI {F_1\rho_\excl2} dx_3...dx_n + \hot,
	\end{eqnarray*}
which is $\log (\FP_\excl2\rho)_1(x_1)$.
As conventional, let $\ve x(t+\delt) \equiv \ve y$ and leave $\ve x$ for
$\ve x(t)$ to avoid confusion. We actually need to find 
	\begin{eqnarray*}
        && \log (\FP_\excl2\rho)_1(y_1) 
		= \log \rho_{1\excl2}(t+\delt; y_1)		\\
	&&\qquad 
	 = \log\rho_1(t; x(t+\delt)) - 
	   \delt \int_{\R^{n-2}} \frac 1 {\rho_{1\excl2}}
		\DI {F_1\rho_\excl2} dx_3...dx_n + \hot		\\
	&&\qquad
	  = \log\rho_1(t; x) + \DI {\log\rho_1}  F_1\delt
	   - \delt \int_{\R^{n-2}} \frac 1 {\rho_{1\excl2}}
		\DI {F_1\rho_\excl2} dx_3...dx_n + \hot.
	\end{eqnarray*}
Taking expectation and multiplying by (-1) on both sides, we obtain	
	\begin{eqnarray*}
	H_{1\excl2}(t+\delt) = H_1(t) 
		- \delt E \parenth{F_1 \DI {\log\rho_1}}
		+ \delt E \int_{\R^{n-2}} \frac 1 {\rho_1}
			\DI {F_1\rho_\excl2} dx_3...dx_n.
	\end{eqnarray*}
So	
	\begin{eqnarray*}
	\dt {H_{1\excl2}} = \lim_{\delt\to0}
		\frac {H_{1\excl2}(t+\delt) - H_1(\delt)} \delt
	=
	     E \int_{\R^{n-2}} \frac 1 {\rho_1}
		\DI {F_1\rho_\excl2} dx_3...dx_n
		-  E \parenth{F_1 \DI {\log\rho_1}}.
	\end{eqnarray*}
On the other hand, from the Liouville equation it is easy to obtain
	\begin{eqnarray}
	\dt {H_1} = \int_{\R^n} \log\rho_1 \DI{F_1\rho} d\ve x.
	\end{eqnarray}
Hence
	\begin{eqnarray}
	&& T_{2\to1} = \dt {H_1} - \dt {H_{1\excl2}}	\cr
	 && \ \ = 
	     \int_{\R^n} \log\rho_1 \DI{F_1\rho} d\ve x
	     - E \int_{\R^{n-2}} \frac 1 {\rho_1}
		\DI {F_1\rho_\excl2} dx_3...dx_n
	     + E \parenth{F_1 \DI {\log\rho_1}}	\cr
	&&\ \ 
	     = -E\bracket{\frac 1 {\rho_1} \int_{\R^{n-2}} 
			\DI{F_1\rho_\excl2} dx_3...dx_n},
	\end{eqnarray}
which is the same as (\ref{eq:T21_det_cont}) in
Theorem~\ref{thm:T21_det_cont}.

\subsection{Properties}

   \begin{thm}
   For a 2D system 
	\begin{eqnarray*}
	\dt{x_1} = F_1(x_1, x_2, t), \\
	\dt{x_2} = F_2(x_1, x_2, t),
	\end{eqnarray*}
   we have
	\begin{eqnarray}
	\dt {H_{1\excl2}} = E\parenth{\DI {F_1}}.
	\end{eqnarray}
   \end{thm}
Remark: This recovers Eq.~(\ref{eq:phys_argu}), the key equation
originally obtained by Liang and Kleeman\cite{LK05} 
through heuristic argument.
Here we rigorously prove it.

   \pf
   When $n=2$, $\rho_\excl2 = \rho_1$, hence
	\begin{eqnarray*}
	\dt{H_{1\excl2}} 
	&=& E \parenth{\frac 1 {\rho_1} \DI {F_1\rho_1} }
	  - E \parenth{\DI {\log\rho_1} F_1} \\
	&=& E\bracket{\DI {F_1} + F_1 \DI {\rho_1} \frac 1 {\rho_1}
		- F_1 \DI {\log\rho_1} } \\
	&=& E\parenth{\DI {F_1}}.
	\end{eqnarray*}
   \qed

   \begin{thm} 	\label{thm:causality_det_cont}
	{\bf Property of causality}\\
	For the system (\ref{eq:gov1})-(\ref{eq:gov3}),
	if $F_1$ is independent of $x_2$, then $T_{2\to1}=0$.
   \end{thm}
   \pf
   If $F_1$ has no dependence on $x_2$, so is $F_1\rho_\excl2$. Thus
   \begin{eqnarray*}
	T_{2\to1}
        &=& -E\bracket{\frac 1 {\rho_1} \int_{\R^{n-2}} 
			\DI{F_1\rho_\excl2} dx_3...dx_n}	\\
	&=& - \int_{\R^n} \rho(x_2|x_1) \DI {F_1\rho_\excl2} d\ve x \\
	&=& - \int_{\R^{n-1}} \DI {F_1\rho_\excl2} dx_1dx_3...dx_n \\
	&=& 0.
   \end{eqnarray*}
where the fact $\int \rho(x_2|x_1) dx_2 = 1$ and the assumption of compact
support have been used.
\qed

\subsection{Application--R\"ossler system}

\begin{figure} [h]
   \begin{center}
   \includegraphics[angle=0,width=0.8\textwidth] {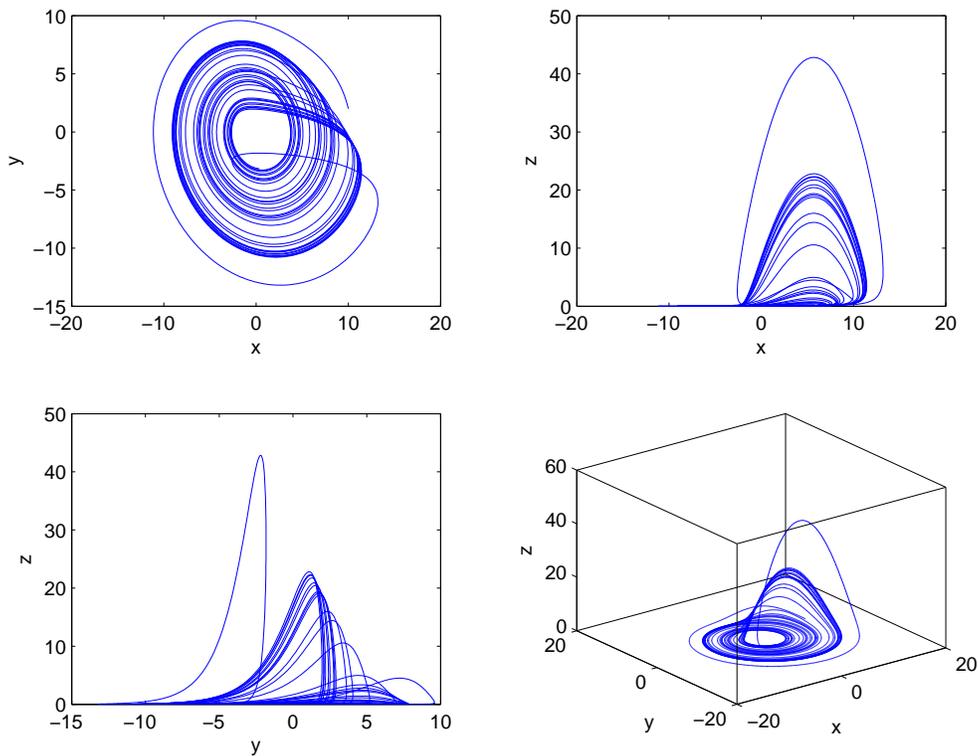}
   \caption
        {The R\"ossler attractor.}
	\protect{\label{fig:rossler}}
   \end{center}
   \end{figure}

In this subsection, we present an application study of 
the information flows within the R\"ossler system: 
	\begin{eqnarray}
	&& \dt x = F_x = -y-z, 		\label{eq:ros1}\\
	&& \dt y = F_y = x + a y,	\label{eq:ros2}\\
	&& \dt z = F_z = b + z(x-c),	\label{eq:ros3}
	\end{eqnarray} 
where $a$, $b$, and $c$ are parameters. Otto E. R\"ossler finds a chaotic
attractor for $a=0.2$, $b=0.2$, $c=5.7$ (\cite{Rossler}), as shown in
Fig.~\ref{fig:rossler}. From the figure the trajectories are limited
within $[-12,12] \times [-14,10] \times [0, 25]$.

\begin{figure} [h]
   \begin{center}
   \includegraphics[angle=0,width=0.5\textwidth] {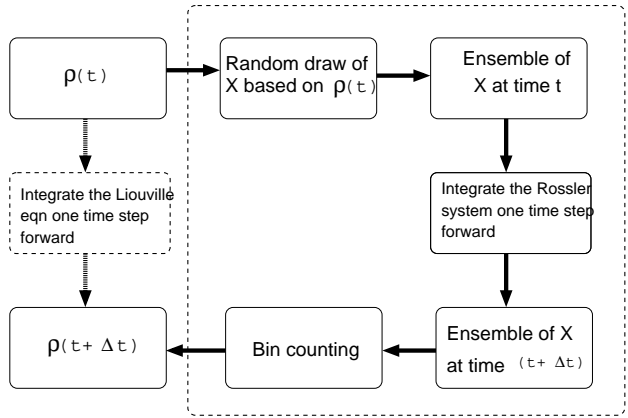}
   \caption
        {A schematic of ensemble prediction. Instead of solving the
         Liouville equation for the density $\rho$, we make random draws
	 according to the initial distribution $\rho(t_0)$ to form an
	 ensemble, then let the R\"ossler system steer forth each member 
 	 of the ensemble. At each time step, bins are counted and the
	 probability density function is accordingly estimated.
	 }
	\protect{\label{fig:ensemble}}
   \end{center}
   \end{figure}

To calculate the information flows, one needs to obtain the joint
probability density function $\rho(x_1,x_2,x_3)$. 
It is, of course, obtainable through solving the Liouville equation
	\begin{eqnarray*}
	\Dt\rho + \frac{\D F_x\rho} {\D x} 
		+ \frac{\D F_y\rho} {\D y}
		+ \frac{\D F_z\rho} {\D z} = 0
	\end{eqnarray*}
with some initial condition $\rho_0$.
However, there is another way, namely, ensemble forecast, which is 
more efficient in terms of computational load. 
As illustrated in Fig.~\ref{fig:ensemble}, 
instead of solving the Liouville equation, we solve the R\"ossler 
systems initialized with an ensemble of initial values of $\ve x$.
This ensemble is formed with entries randomly drawn according to
the initial pdf $\rho_0$.
At each time step, we count the bins thus obtained and estimate
the pdf. The resulting pdf is the desired $\rho$. 

The R\"ossler system
(\ref{eq:ros1})-(\ref{eq:ros2}) is solved using the second order 
Runge-Kutta method with a time step size $\delt=0.01$. 
A typical computed trajectory is plotted in Fig.~\ref{fig:rossler}. 
The initial conditions are randomly drawn according to 
a Gaussian distribution $N(\veg\mu, \vveg\Sigma)$, 
the mean vector and covariance matrix being, respectively,
	\begin{eqnarray*}
	\veg\mu = \left[\begin{array}{l}
			8	\\
			2	\\
			10	
			\end{array}\right],
	\qquad
	\vveg\Sigma = \left[\begin{array}{lll}
			4 & 0 	    & 0  \\
			0 	 & 4 & 0 \\
			0 	 &  0 	    & 4 
		     \end{array}\right].
	\end{eqnarray*}
The initial mean values are chosen rather randomly
(in reference to Fig.~\ref{fig:rossler});
$\mu_x$ is chosen large to make
	$\dt H = E(\nabla\cdot\ve F) = \mu_x - 5.5$
positive.

Pick a computation domain 
	$\Omega \equiv [-16,16] \times [-18,14] \times [-4, 28]$,
clearly covers the attractor. We discretize it into
$320\times320\times320 = 32,768,000$~bins with 
$\Delta x = \Delta y = \Delta z = 0.1$. 
To ensure one draw for each bin on average,
in the beginning we make $32,768,000$ random draws.
As the ensemble scheme is carried forth, 
$\rho$ and all other statistics can be estimated as a function of time.
By Theorem~\ref{thm:T21_det_cont} the information flow rates are 
computed accordingly.


\begin{figure} [h]
   \begin{center}
   \includegraphics[angle=0,width=0.5\textwidth] {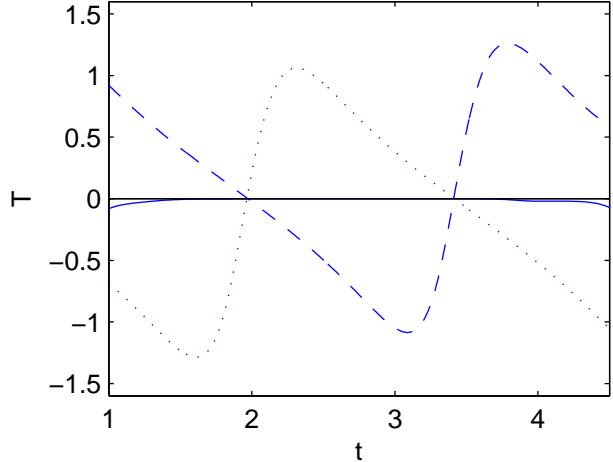}
   \caption
	{The time series of the
	information flow rates within the R\"ossler system 
	(in nats per unit time). 
	Dashed: $T_{y\to x}$; dotted: $T_{x\to y}$; solid: $T_{z\to x}$.
	Other flows are essentially zero in this duration.
	The initial segments are not shown as some trajectories 
	are still outside the attractor.}
	\protect{\label{fig:rossler_info}}
   \end{center}
   \end{figure}

For a system with three components $(x,y,z)$, there are in total 6 flow
pairs: $T_{x\to y}$, $T_{y\to z}$, $T_{z\to x}$, $T_{x\to z}$, $T_{y\to
x}$, $T_{x\to z}$. A first examination of the system tells that 
$dy/dt$ does not depend on $z$ and $dz/dt$ does not depend on $y$.
By the property of causality (Theorem~\ref{thm:causality_det_cont}),
$T_{z\to y}$ and $T_{y\to z}$ must vanish.
The computational results reconfirm this. In Fig.~\ref{fig:rossler_info},
the two are essentially zero. What makes the results surprisingly
interesting is that $T_{x\to z}$ is also insignificant, 
while the dependence of $dz/dt$ on $x$ is explicitly specified.
Besides, $T_{z\to x}$ is also small. In the figure are essentially 
the flows between $x$ and $y$: $T_{y\to x}$ and $T_{x\to y}$.

   \begin{figure} [h]
   \begin{center}
   \includegraphics[angle=0,width=0.5\textwidth] {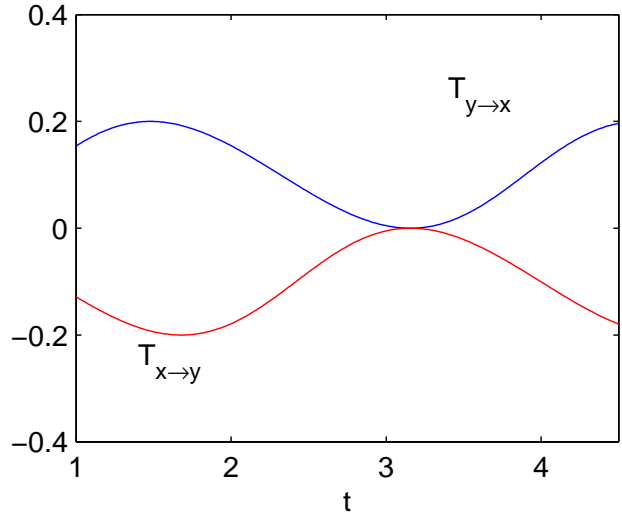}
   \caption
	{The time series of the
	information flow rates within the amplifying harmonic system 
	as shown in the text (in nats per unit time). 
	}
	\protect{\label{fig:rossler_linear}}
   \end{center}
   \end{figure}

The above information flow scenario motivates us to check the system with
only $x$ and $y$ two components. This is an amplifying harmonic oscillator
	$\dt {\ve x} = \vve A \ve x$ where $\vve A = \mat 0 {-1} 1 a$,
a linear system allowing the information flow, say, from $y$ to $x$, to be
simply expressed as $T_{y\to x} = a_{12} \frac{\sigma_{12}}{\sigma_{11}}$
(see below in section~\ref{sect:linear}).
That is to say, here the covariance matrix $\vve\Sigma = (\sigma_{ij})$ 
completely determines the flow.
The evolution of $\vve\Sigma$ follows 
	$$\dt\Sigma = \vve A \vve\Sigma + \vve\Sigma\vve A^T.$$
Initialized by $\mat 4 0 0 4$, $\sigma_{ij}$ can be easily computed;
the resulting $T_{y\to x}$ and $T_{x\to y}$ are shown in
Fig.~\ref{fig:rossler_linear}. Comparing to those in 
Fig.~\ref{fig:rossler_info}, the general trend, including the period, 
seems to be similar, 
though the geometry of the curves has been modified from harmonic into a
seesaw ones.
Besides, the $T_{x\to y}$ ($T_{y\to x}$) is always negative (positive)
for the harmonic oscillator, while for the R\"ossler system, they 
can be both negative and positive.
Note the parameter $a$ in $\vve A$ does not explicitly appear in the
formula, but it does contributes to the generation of the information flow.
One may easily check that, if it is zero, then $\dt\Sigma = 0$, and hence
the flow rates will stay zero if originally $\sigma_{12}=0$.

The above example is just used for the demonstration of application and, 
in some cases, for the validation of the proven theorems such as the 
property of causality. The seemingly vanishing $T_{x\to z}$ in spite of 
the dependence of $dz/dx$ on $x$ for sure deserves further investigation
but is beyond the scope of this study. Here we just want to mention that
this does conform to the observations with complex systems--Emergence does
not result from rules only (e.g., \cite{Crutchfield}-\cite{Goldstein}).
It has long been found that regular patterns may emerge out of 
irregular motions with some simple preset rules;
a good example is the 2D turbulent flow in natural world (e.g.,
\cite{McWilliams}).
Clearly, these simple, rudimentary rules are not enough for explaining the
causal efficacy and the bottom-up flow of information that leads to the
emergence of the organized structure.
As commented by Corning\cite{Corning},
``Rules, or laws, have no causal efficacy; 
they do not in fact generate anything... the
underlying causal agencies must be separately specified.''
We shall see a more remarkable example in the following subsection.

\subsection{Application---The truncated Burgers-Hopf system revisited}

Here we re-examine the Truncated Burgers-Hopf system (TBS hereafter), a
chaotic system which seemingly has rather simple information flow structures 
in the studies of Liang and Kleeman\cite{LK07b}.
For a detailed description of the system itself, 
see \cite{Abramov}. In this section we only
examine the following particular case:
	\begin{eqnarray}	
&&\dt {x_1} = F_1(\ve x) = x_1 x_4 - x_3 x_2, \label{eq:burgersa} \\
&&\dt {x_2} = F_2(\ve x) = -x_1 x_3 - x_2 x_4, \label{eq:burgersb}	\\
&&\dt {x_3} = F_3(\ve x) = 2 x_1 x_2,	\label{eq:burgersc} \\
&&\dt {x_4} = F_4(\ve x) = -x_1^2 + x_2^2.	\label{eq:burgersd}
	\end{eqnarray}
As we have described before, 
the system is intrinsically chaotic, with
a strange attractor embedded in
	$$[-24.8, 24.6] \times [-25.0, 24.5] \times [-22.3, 21.9]
	  \times [-23.7, 23.7].$$

The information flow within the TBS cannot be found analytically.
As before, we use the ensemble prediction technique to estimate
the density evolution, and then evaluate the $T$'s. 
The setting and procedure are made precisely the same as that in \cite{LK07b}
in order to facilitate a comparison.
Details are referred to the original paper and will not be presented 
here.

\begin{figure} [h]
   \begin{center}
   \includegraphics[angle=0,width=1\textwidth] {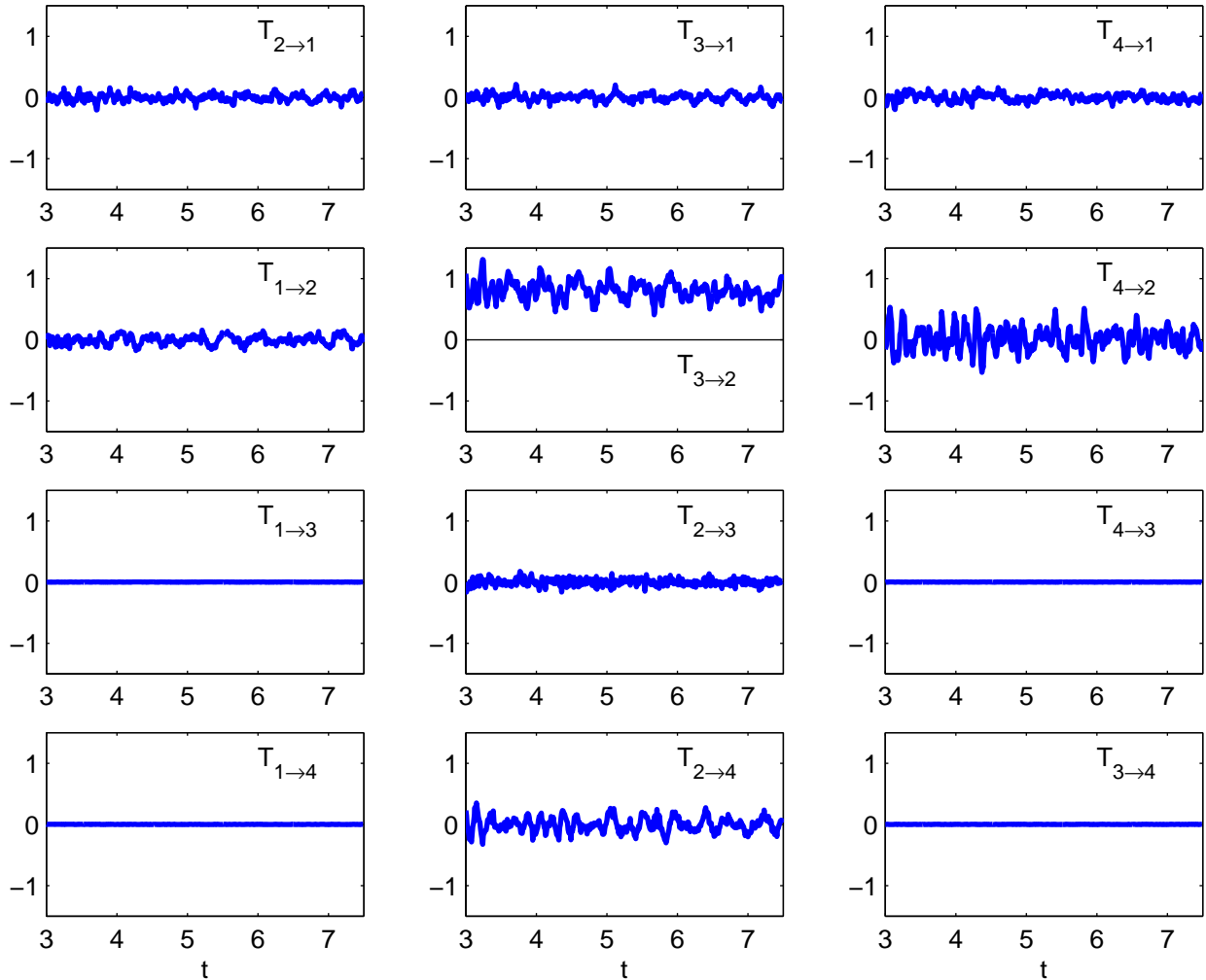}
   \caption
        {Information flows between the components of the 4D truncated 
	Burgers-Hopf system in the invariant chaotic attractor.}
	\protect{\label{fig:Burgers_transfer}}
   \end{center}
   \end{figure}

Figure~\ref{fig:Burgers_transfer} plots the results for the case with
a Gaussian initial distribution $N(\veg\mu, \vveg\Sigma)$, where
	\begin{eqnarray*}
	\veg\mu = \left[\begin{array}{l}
			\mu_1	\\
			\mu_2	\\
			\mu_3	\\
			\mu_4
			\end{array}\right],
	\qquad
	\vveg\Sigma = \left[\begin{array}{llll}
			\sigma_1^2 & 0 	    & 0  & 0\\
			0 	 & \sigma_2^2 & 0 & 0\\
			0 	 &  0 	    & \sigma_3^2 &0	\\
			0 	 &  0 	    &  0         & \sigma_4^2
		     \end{array}\right],
	\end{eqnarray*}
with $\mu=(9,9,9,9)$, $\sigma^2=(9,9,9,9)$. 
Shown specifically are the time rates of the 12 information flows:
	\begin{eqnarray*}
	&& T_{2\to1},	\quad T_{3\to1},  \qquad T_{4\to1}; \\
	&& T_{1\to2},	\quad T_{3\to2},  \qquad T_{4\to2}; \\
	&& T_{1\to3},	\quad T_{2\to3},  \qquad T_{4\to3}; \\
	&& T_{1\to4},	\quad T_{2\to4},  \qquad T_{3\to4}. 
	\end{eqnarray*}
The results are qualitatively the same as before in \cite{LK07b}.
That is to say, except for $T_{3\to2}$, which is distinctly different from
zero, all others are either negligible, or oscillatory around zero.
But, of course, the present flows are much smaller in magnitude, in
comparison to the one obtained before in \cite{LK07b} using the approximate
formula.

\section{Stochastic mapping}	\label{sect:stoch_map}

\subsection{Derivation}

Consider the system
	\begin{eqnarray}	\label{eq:stoch_discrete_sys}
	\ve x(\tau+1) = \mapping(\ve x(\tau)) + \vve B(\ve x) \ve w,
	\end{eqnarray}
where $\mapping: \R^n\to\R^n$ 
is an $n$-dimensional mapping, $\vve B$ is an $n\times m$ matrix, 
and $\ve w$ an $m$-dimensional normally distributed random vector,
representing an $m$-dimensional standard Wiener process. 
Without loss of generality, we assume the covariance matrix of 
$\ve w$, $\vve \Sigma = \vve I$, since the perturbation amplitude can be put 
into $\vve B$.

In general, $\vve B$ may depend on $\ve x$. But this complicates the
derivation a lot. For simplicity, in this section 
we only consider the case when $\vve B$ 
is a constant $n\times m$ matrix.  As $\ve x(\tau)$ is taken to $\ve
x(\tau+1)$, there exists an operator, written $\FP: L^1(\R^n) \to
L^1(\R^n)$, steering the pdf at time step $\tau$, $\rho$, to the pdf at step
$\tau+1$, $\FP\rho$. Since $\ve x(\tau)$ and $\ve w$ are independent, 
if $\vve B$ is a constant matrix, one may view $\ve x(\tau+1)$ as
the sum of two independent random variables, and then conjecture that
$\FP\rho$ be the convolution of $\FP_\mapping\rho$ and some joint Gaussian
distribution. Here $\FP_\mapping$ stands for the F-P operator associated
with the mapping $\mapping$. This is indeed true, as is stated in the
following theorem.

   \begin{thm}
   	\begin{eqnarray}
   	\FP\rho(\ve y) = \int_{\R^n} \FP_\mapping 
	\rho(\ve y - \vve B\ve w)
	\cdot \rho_w(\ve w) d\ve w
   	\end{eqnarray}
   where 
	\begin{eqnarray*}
	\rho_w(\ve w) = (2\pi)^{-m/2} \parenth{\det\vve\Sigma}^{-1/2}
		e^{-\frac12 \ve w^T {\vve\Sigma}^{-1} \ve w}.
	\end{eqnarray*}
   \end{thm}
   \pf
    We first assume that $\mapping$ is invertible to make the approach more
    transparent to the reader. As always, write $\ve x(\tau+1)$ as 
    $\ve y$ to avoid confusion. Make a transformation:
	\begin{eqnarray}
	\Pi:  \left\{\begin{array}{l}
		     \ve y = \mapping(\ve x) + \vve B \ve w,\\
		     \ve z = \ve w.
		     \end{array}
		\right.
	\end{eqnarray}
Its Jacobian 
	\begin{eqnarray}
	J_\pi = \det \Jacobian {\ve y, \ve z} {\ve x, \ve w}
	      = \det \mat {\Dx\mapping} {\vve B} {\vve 0}  {\vve I}
	      = \det \parenth{\frac {\D\mapping} {\D \ve x}}
	      = J_\mapping \equiv J.
	\end{eqnarray}
The inverse mapping is
	\begin{eqnarray}
	\Pi^{-1}:  \left\{\begin{array}{l}
		     \ve x = \inmapping(\ve y - \vve B \ve z),\\
		     \ve w = \ve z.
		     \end{array}
		\right.
	\end{eqnarray}
For any $S_y \in \R^n$, $S_z \in \R^m$,
	\begin{eqnarray*}
	\int_{S_y \times S_z} \rho_{yz}(\ve y, \ve z) d\ve y d\ve z
	&=& \int_{\Pi^{-1}(S_y\times S_z)} 
		\rho_{xw} (\ve x, \ve w) d\ve x d\ve w	\\
	&=& \int_{S_y \times S_z} \rho_{xw}
		\parenth{\Pi^{-1}(\ve y, \ve z)} \cdot \abs{J_\pi^{-1}}
		d\ve y d\ve z.
	\end{eqnarray*}
So 
	\begin{eqnarray*}
	\rho_{yz}(\ve y, \ve z)
	&=& \rho_{xw}
	    \parenth{\Pi^{-1}(\ve y, \ve z)} \cdot \abs{J_\pi^{-1}} \\
	&=& \rho_{xw}
	    \parenth{\inmapping(\ve y - \vve B\ve z), \ve z)} 
			\cdot \abs{J^{-1}} \\
	&=& \rho\parenth{\inmapping(\ve y - \vve B\ve z)}
			\cdot \abs{J^{-1}} 
		\cdot \rho_w(\ve z),
	\end{eqnarray*}
where the independence between $\ve x$ and $\ve w$ has been used
(hence $\rho_{xw} = \rho_x \cdot \rho_w$).
$\FP(\ve y) = \rho_y(\ve y)$ is thence the marginal density by integrating
out $\ve z$:
	\begin{eqnarray*}
   	\FP\rho(\ve y) = \int_{\R^n} 
	 \rho\parenth{\inmapping(\ve y - \vve B\ve z)}
			\cdot \abs{J^{-1}} 
		\cdot \rho_w(\ve z)\ d\ve z.
	\end{eqnarray*}
Since 
   $\rho\parenth{\inmapping(\ve y)} \cdot \abs{J^{-1}} 
    = \FP_\mapping\rho(\ve y)$,
the theorem thus follows.

When $\mapping$ is singular or noninvertible, let its F-P operator be
$\FP_\mapping$, then $\forall S_y \in \R^n$, $S_z \in \R^m$,
	\begin{eqnarray*}
	&& 
	\int_{S_y\times S_z} \rho_{yz} (\ve y, \ve z) d\ve y d\ve z
	= \int_{\Pi^{-1}(S_y\times S_z)} \rho_{xw}(\ve x, \ve z) 
			d\ve x d\ve z	\\
	&& = \int_{\Pi^{-1}(S_y\times S_z)} \rho(\ve x) \cdot \rho_w(\ve z)
			d\ve x d\ve z	\\
	&& = \int_{S_z} \rho_w(\ve z) d\ve z
	  \int_{\inmapping S_{y-Bz}} \rho(\ve x) d\ve x. \\
	&& = \int_{S_z} \rho_w(\ve z) d\ve z
		\int_{S_{y-Bz}} \FP_\mapping \rho(\ve x)\ d\ve x \\
	&& = \int_{S_z} \rho_w(\ve z) d\ve z
		\int_{S_y} \FP_\mapping(\ve y - \vve B\ve z) d\ve y.
	\end{eqnarray*}
The conclusion follows accordingly.
\qed

With the above theorem, the information flow can be easily computed.
Note the theorem actually states that 
	\begin{eqnarray}
	\FP\rho(\ve y) = E_w \FP_\mapping\rho(\ve y - \vve B \ve w)
	\end{eqnarray}
where $E_w$ signifies the expectation taken with respect to $\ve w$.
So 
	\begin{eqnarray*}
	H_1(\tau+1) &=& - E_y \log(\FP\rho)_1 (y_1)	\\
	    &=& - E_x \bracket{\log \int_{\R^{n-1}} 
	            E_w \FP_\mapping\rho(\ve y - \vve B \ve w) dy_2...dy_n}\\
	    &=& - E_x \bracket{\log E_w (\FP_\mapping\rho)_1
				      (y_1 - \ve B_1 \ve w) },
	\end{eqnarray*}
where $\ve B_1 \equiv (b_{11},\ b_{12}, ..., b_{1m})$ is the row vector.
Likewise,
	\begin{eqnarray}
	\FP_\excl2\rho(\ve y_\excl2) 
	= E_w \FP_{\mapping_\excl2} \rho(\ve y_\excl2 - \vve B_\excl2 \ve w).
	\end{eqnarray}
In the equation, the subscript $\excl2$ in the vector(s) and matrix
means the second row is removed from the corresponding entities.
So 
	\begin{eqnarray*}
	H_{1\excl2}(\tau+1) 
	    &=& - E_y \log(\FP_\excl2\rho)_1 (y_1)	\\
	    &=& - E_x \bracket{\log \int_{\R^{n-2}} 
	            E_w \FP_{\mapping_\excl2}
		\rho(\ve y_\excl2 - \vve B_\excl2 \ve w) dy_3...dy_n}\\
	    &=& - E_x \bracket{\log E_w (\FP_{\mapping_\excl2} \rho)_1
				      (y_1 - \ve B_1 \ve w) }.
	\end{eqnarray*}
Subtract $H_{1\excl2}(\tau+1)$ from $H_1(\tau+1)$, and the information
flow $T_{2\to1}$ follows:
	\begin{thm}
	\begin{eqnarray}
	T_{2\to1} = 
	    E_x \bracket{\log E_w (\FP_{\mapping_\excl2} \rho)_1
				      (y_1 - \ve B_1 \ve w) }
	    - E_x \bracket{\log E_w (\FP_\mapping\rho)_1
				      (y_1 - \ve B_1 \ve w) }.
	\end{eqnarray}	
	\end{thm}

\subsection{Properties}
	\begin{thm}
	{\bf Property of causality}\\
	For the system (\ref{eq:stoch_discrete_sys}), if $\mapping_1$
	and $\ve B_1$ are independent of $x_2$, then  $T_{2\to1} = 0$.
	\end{thm}
\pf 
As we proved for the deterministic case, if $\mapping_1$ is independent
of $x_2$, then 
	$(\FP_\mapping\rho)_1 \overset{a.e.}= (\FP_{\mapping_\excl2}\rho)_1$.
If further $\ve B_1$ has no dependence on $x_2$, then the above
$H_1(\tau+1)$ and $H_{1\excl2}(\tau+1)$ are equal, and hence 
$T_{2\to1} = 0$. \qed

\subsection{Application: A noisy H\'enon\ map}

We now reconsider the benchmark systems that have been examined before, but
with Gaussian noise added. The baker transformation is not appropriate here, 
since the added noise perturbation will take $\ve x$ outside the domain
$[0,1]$. We hence only look at the H\'enon map 
	$\mapping: \R^2 \to \R^2:$
	\begin{eqnarray}
	\left\{\begin{array}{l}
	\mapping_1(x_1,x_2) = 1 + x_2 - \alpha x_1^2,\\
	\mapping_2(x_1,x_2) = \beta x_1,
	\end{array}\right.
	\end{eqnarray}
with parameters $\alpha, \beta > 0$, and consider only the case 
$T_{1\to2}$  which has been shown as a benchmark case.
Now perturb $\mapping$ to make a
stochastic mapping:
	\begin{eqnarray}
	\ve x(\tau+1) = \mapping(\ve x(\tau)) + \vve B \ve w
	\end{eqnarray}
where $\vve B = (b_{ij})$ is a constant matrix, $\ve w \sim N(0,\vve I)$.
Let $\ve B_i \equiv (b_{i1}, b_{i2})$ denote a row vector. It is easy to
see that $\mapping$ is invertible; in fact, 
	$J = \mat {-2\alpha^2} 1 \beta 0 = -\beta \ne 0$.
The inverse is
	\begin{eqnarray}
	\inmapping(x_1,x_2) = \parenth{\frac {x_2} \beta,\ 
			x_1 - 1 + \frac \alpha {\beta^2} x_2^2 }.
	\end{eqnarray}
Thus 
	\begin{eqnarray}
	\FP_\mapping\rho(x_1,x_2) = \rho(\inmapping(x_1,x_2)) \abs{J^{-1}}
	= \rho\parenth{\frac {x_2} \beta,\ 
			     x_1 - 1 + \frac\alpha {\beta^2} x_2^2}
		\cdot \beta^{-1}
	\end{eqnarray}
So $\FP\rho(\ve y)= E_w \FP_\mapping\rho(\ve y - \vve B\ve w)$,
and
	\begin{eqnarray*}
	(\FP\rho)_2(y_2) 
	&=& \int_\R dy_1 E_w \frac 1 \beta 
		    \rho\parenth{\frac {y_2 - \ve B_2\ve w} \beta,\  
		    y_1 - \ve B_1\ve w - 1 + \frac\alpha{\beta^2}
			(y_2 - \ve B_2 \ve w)^2} \\
	&=& \frac 1 \beta E_w \rho_1\parenth{\frac {y_2 - \ve B_2\ve w} \beta}.
	\end{eqnarray*}

If $x_1$ is frozen, $\mapping_2(x_1,x_2) = \beta x_1$ is a constant. 
Hence $H_{2\excl1}(\tau+1) = 0$, and
	\begin{eqnarray}
	T_{1\to2} 
	 &=& H_2(\tau+1) - H_{2\excl1}(\tau+1)	\cr
	 &=& -E\bracket{\log \frac 1 \beta E_w
			\rho_1\parenth{\frac {y_2-\ve B_2\ve w} \beta} }
	   - 0		\cr
	 &=& \log\beta - E_w E_x \log 
		\rho_1 \parenth{\frac {y_2-\ve B_2\ve w} \beta}   \cr
	 &=& \log\beta - E_w E_x \log 
		\rho_1 \parenth{x_1 - \frac {\ve B_2\ve w} \beta} \cr
	 &=& \log\beta + {\mathscr F} H_1.
	\end{eqnarray}
Here ${\mathscr F} H_1$ is the functional $H_1$ applied by a Gaussian filter.
One may understand it as $H_1$ smeared out by a Gaussian filter. It is less
than $H_1$, so the noise addition makes the system lose some information,
compared to $T_{1\to2} = \log\beta + H_1$ in the deterministic case.

\section{Continuous-time stochastic systems}	\label{sect:stoch_flow}

\subsection{Derivation}

Following what we have done in section~\ref{sect:det_flow}, we derive the
information flow within a continuous-time stochastic system by taking the
limit of the corresponding discrete stochastic mapping. In doing this, the
results in the preceding section are ready for use. But, as noted,
in the above derivation we have assumed a constant matrix $\vve B$, 
a simplified case allowing for a clear expression of information flow.
(This case does have realistic relevance, though.) 
For a time continuous system, this assumption actually can be completely 
relaxed. In the following we will see why.

Consider a system
	\begin{eqnarray}	\label{eq:stoch_gov}
	d\ve x= \ve F(t; \ve x) dt + \vve B(t; \ve x) d\ve w,
	\end{eqnarray}
where $\ve x$ and $\ve F$ are $n$-dimensional vector, $\vve B$ is an
$n\times m$ matrix, and $\ve w$ an $m$-vector of standard Wiener process.
Note that $\vve B$ can be a function of both $\ve x$ and time $t$.
This above equation may also be written as
	\begin{eqnarray}
	\dt{\ve x} = \ve F(t; \ve x) + \vve B(t; \ve x) \dot{\ve w},
	\end{eqnarray}
where $\dot{\ve w}$ a vector of white noise, or, in component form, 
	\begin{eqnarray}
	&&\dt {x_1} = F_1(t; \ve x) + {\ve B}_1(t; \ve x) \dot{\ve w},\\
	&&\dt {x_2} = F_2(t; \ve x) + {\ve B}_2(t; \ve x) \dot{\ve w},\\
	&&\quad\vdots 	\qquad\qquad\quad\vdots 		      \\
	&&\dt {x_n} = F_n(t; \ve x) + {\ve B}_n(t; \ve x) \dot{\ve w}.
	\end{eqnarray}
In the equations we have used ${\ve B}_i$ to indicate the 
$i$-th row vector of $\vve B$.
Now consider (\ref{eq:stoch_gov})
on a small interval $[t, t+\delt]$. 
Euler-Bernstein differencing,
	\begin{eqnarray}
	\ve x(t+\delt) = \ve x(t) + \ve F\delt + \vve B \delw.
	\end{eqnarray}
This motivates the introduction of a transformation
	\begin{eqnarray}
	\Pi: \left\{\begin{array}{l}
	     \ve y = \ve x + \ve F(\ve x) \delt + \vve B(\ve  x)\delw, \\
	     \ve z = \delw.
	     \end{array}\right.
	\end{eqnarray}
As shown in the discrete mapping case, generally this transformation cannot
be inverted. But for this special case where $\delt$ and $\delw$ are small,
the inversion can be done asymptotically. In fact,
	\begin{eqnarray*}
	\ve y 
	&=& \ve x + [\ve F(\ve y) + \hot] \delt
	      + [\vve B(\ve y) \delw + \nabla (\vve B(\ve y) \delw) 
		(\ve x - \ve y)]	+  o(\Delta w^2)	\\
	&=& \ve x + \ve F(\ve y) \delt + \vve B(\ve y) \delw
	  	+ [\nabla(\vve B\delw)] (-\ve F\delt - \vve B\delw) 
		+ \hot.
	\end{eqnarray*}
Note, here the higher order terms means terms with order higher than
$\delt$ or $(\Delta w)^2$ --- We will see soon that $E (\Delta w)^2 =
\delt$. The above expansion helps invert $\Pi$ to
	\begin{eqnarray}	\label{eq:inv_pi}
	\Pi^{-1}: \left\{\begin{array}{l}
	     \ve x = \ve y - \ve F \delt - \vve B \ve z 
		     + \nabla(\vve B\ve z) (\vve B\ve z) + \hot, \\
	     \delw = \ve z.
	     \end{array}\right.
	\end{eqnarray}

The following proposition finds the Jacobian associated with the inverse
transformation. 
   \begin{prop}
    Define the double dot of two dyadics $\vve A$ and $\vve B$ as
	$\vve A : \vve B = \sum_{i,j} a_{ij} b_{ji}$, 
    then
	\begin{eqnarray}
	J^{-1} = 1 - \nabla\cdot\ve F\delt - \nabla\cdot(\vve B\ve z)
		+ \frac12 \nabla\nabla : (\vve B\ve z \ve z^T \vve B^T)
		+ \hot.
	\end{eqnarray}
   \end{prop}
\pf
By definition
	\begin{eqnarray}		\label{eq:J_inv}
	J^{-1} = \Jacobian {\ve x, \delw} {\ve y, \ve z}
	       = \det\mat {\DD {\ve x} {\ve y}} {\DD {\ve x} {\ve z}}
			  {\DD \delw {\ve y}}   {\DD \delw {\ve z}}
	       = \det\mat {\ \DD {\ve x\ } {\ve y}} {-\vve B + ...}
			  {\ \ \vve 0}              {\quad\ \vve I}
	       = \det \parenth{\DD {\ve x} {\ve y}}.
	\end{eqnarray}
The key is the evaluation of $\det \parenth{\DD {\ve x}{\ve y}}$.
By (\ref{eq:inv_pi}), it is, up to $\hot$, the determinant of
	\begin{eqnarray}
	\matthree 
	{1-\DyI {F_1}\delt -\sum_k \DyI {b_{1k}}  z_k
	     + \DyI\ \parenth{\sum_{l,k,s}
		\DD {b_{1k}}{y_l} z_k b_{ls} z_s }}
			{...}	 
	{-\Dyn {F_1}\delt - \sum_k \Dyn{b_{1k}} z_k 
	     + \Dyn\ \parenth{\sum_{l,k,s}
		\DD {b_{1k}}{y_l} z_k b_{ls} z_s }}
		      \vdots    \ddots		\vdots
	{-\DyI {F_n}\delt - \sum_k \DyI{b_{nk}} z_k 
	     + \DyI\ \parenth{\sum_{l,k,s}
		\DD {b_{nk}}{y_l} z_k b_{ls} z_s }}
		       	{...}	
	{1-\Dyn {F_n}\delt - \sum_k \Dyn {b_{nk}} z_k
	     + \Dyn\ \parenth{\sum_{l,k,s}
		\DD {b_{nk}}{y_l} z_k b_{ls} z_s }}
	\end{eqnarray}
Recall that, for an $n\times n$ matrix $\vve A = (a_{ij})$, 
	\begin{eqnarray}
	\det\vve A = \sum_{\sigma\in P_n} \sgn(\sigma)
		\prod_{i=1}^n a_{i,\sigma_i}
	\end{eqnarray}
where $P_n$ is the totality of permutations of $\{1,2,...,n\}$.
By this formula, the terms of order $\delt$ and $\Delta w$ are easy to find;
they can only come from the diagonal entries. For terms of order $(\Delta
w)^2$, there are three sources:
	\begin{itemize}
	\item[(1)] the last term at each diagonal entry, 
		together with $n-1$ 1's;
	\item[(2)] multiplication of two entries at $(i,i)$ and 
		$(j,j)$, $i\ne j$, together with $n-2$ 1's on the diagonal;
	\item[(3)] similar to (2), but with entries at 
		$(i,j)$ and $(j,i)$, $i\ne j$. 
	\end{itemize}
In (3) the order between $i$ and $j$ switches and it has 
$\sgn=-1$ for its permutation. Except for (3) involving off-diagonal
entries, all others are from the diagonal. So
	\begin{eqnarray*}
	&&
	\det\parenth{\DD{\ve x}{\ve y}}
	= \prod_{i=1}^n
	   \bracket
        	{1-\DD {F_i} {y_i} \delt -\sum_k \DD {b_{ik}} {y_i}  z_k
	     + \sum_{l=1}^n \sum_{k,s=1}^m \DD{\ }{y_i}
		\parenth{b_{ls} \DD {b_{ik}}{y_l}} z_k z_s }	\\
	&& \qquad\qquad
	+ \frac12 \sum_{\overset{i,j}{i\ne j}} (-1)^1
		\parenth{-\sum_k \DD {b_{ik}}{y_j} z_k} 
		\parenth{-\sum_s \DD {b_{js}} {y_i} z_s}
	+ \hot.
	\end{eqnarray*}
Notice the factor $\frac12$ in the last term. Because of the symmetry
between $i$ and $j$ and they repeat once when summed over $i,j=1,n$.
Thus
	\begin{eqnarray*}
	&&
	\det\parenth{\DD{\ve x}{\ve y}}
	= 1 - \sum_i \DD {F_i} {y_i} \delt
	    - \sum_i \sum_k \DD {b_{ik}} {y_i} z_k	\\
	&&\qquad
	    + \frac12 \sum_{i\ne j} \sum_k \Dyi {b_{ik}} z_k \cdot
				    \sum_s \Dyj {b_{js}} z_s
	    + \sum_{i,j} \sum_{k,s} \Dyi\ \parenth{b_{js} 
			\Dyj{b_{ik}}} z_k z_s 		\\
	&&\qquad
	- \frac12 \sum_{i\ne j} 
		\sum_k \DD {b_{ik}}{y_j} z_k  \cdot
		\sum_s \DD {b_{js}} {y_i} z_s
	+ \hot.
	\end{eqnarray*}
Notice
	\begin{eqnarray*}
	\DyiDyj {b_{ik} b_{js}} = 
	\Dyi {b_{ik}} \Dyj {b_{js}} + b_{ik} \DyiDyj {b_{js}}
	+ \Dyi {b_{js}} \Dyj {b_{ik}} + b_{js} \DyiDyj {b_{ik}},
	\end{eqnarray*}
and
	\begin{eqnarray*}
	&&
	\sum_{k,s}\sum_{i,j} \Dyi b_{js} \Dyj{b_{ik}} z_k z_s
	= \sum_{k,s} \sum_{i,j} \Dyi {b_{js}} \Dyj{b_{ik}} z_k z_s
	  + \sum_{k,s} \sum_{i,j} b_{js} \DyiDyj {b_{ik}} z_k z_s \\
	&&\ \
	= 
	\frac12 \sum_{k,s} \sum_{i\ne j} \Dyi {b_[js} \Dyj {b_{ik}} z_k z_s
	  + \frac12 \sum_{k,s} \sum_i \Dyi {b_{is}} \Dyi {b_{ik}} z_k z_s
	  + \frac12 \sum_{k,s} \sum_{i,j} \Dyi {b_{js}} \Dyj{b_{ik}} z_k z_s\\
	&&\ \ 
	  + \frac12 \sum_{k,s}\sum_{i,j} \parenth{
		b_{js} \DyiDyj {b_{ik}} + b_{ik} \DyiDyj {b_{js}}} 
		z_k z_s.
	\end{eqnarray*}
The last parenthesis holds because the two pairs $(i,k)$ and $(j,s)$
may be switched under the summation without changing the result.
Thence
	\begin{eqnarray*}
	&&
	\det\parenth {\DD {\ve x} {\ve y}}
	= 1 - \sum_i \Dyi {F_i} \delt - \sum_i\sum_k \Dyi {b_{ik}} z_k 
	 + \frac12 \sum_{i,j} \sum_{k,s} \Dyi {b_{ik}} \Dyj {b_{js}} z_k z_s\\
	&&\ \ \ \ 
	 + \frac12 \sum_{i,j} \sum_{k,s} \Dyi {b_{js}} \Dyj {b_{ik}} z_k z_s
	 + \frac12 \sum_{i,j} \sum_{k,s} \parenth{
		b_{js} \DyiDyj {b_{ik}} + b_{ik} \DyiDyj {b_{js}}
						} z_k z_s	
	  + \hot	\\
	&&\ \ 
	= 1 - \sum_i \Dyi {F_i} \delt - \sum_i\sum_k \Dyi {b_{ik}} z_k
	    + \frac12 \sum_{i,j} \DyiDyj {\sum_{k,s} b_{ik} z_k z_s b_{js}}
	   + \hot \\
	&&\ \
	= 1 - \nabla\cdot\ve F \delt - \nabla\cdot (\vve B \ve z)
	    + \frac12 \nabla\nabla : (\vve B \ve z\ve z^T \vve B^T)
	    + \hot,
	\end{eqnarray*}
which is $J^{-1}$ by (\ref{eq:J_inv}).
\qed

With $J^{-1}$, we can then evaluate the operator $\FP$ and hence arrive  at
$\dt {H_1}$ and $\dt {H_{1\excl2}}$.
   \begin{prop}		
    Let $\vve B \vve B^T \equiv \vve G = (g_{ij})$. The time rate of change of
    $H_1$ is
	\begin{eqnarray} \label{eq:dH1}
	\dt {H_1} = -E\bracket{F_1 \DI {\log\rho_1}}
		  - \frac12 E\bracket{g_{11} \DIDI {\log\rho_1}}.
	\end{eqnarray}
   \end{prop}
\pf
For any subset $S_y \in \R^n$, $S_z \in \R^m$,
	\begin{eqnarray*}
	&&
	\int_{S_y\times S_z} \rho_{yz} (\ve y, \ve z) d\ve y d\ve z
	= \int_{\Pi^{-1}(S_y\times S_z)} \rho_{xw} (\ve x, \delw) 
		d\ve x d delw	\\
	&&\qquad
	= \int_{S_y\times S_z} \rho_{xw}(\ve y - \ve F\delt - \vve B\ve z
		+ \nabla(\vve B\ve z) \cdot (\vve B\ve z),\ \ve z)
		\cdot \abs{J^{-1}}\ d\ve y d\ve z	\\
	&&\qquad
	= \int_{S_y\times S_z}
	  \rho(\ve y - \ve F\delt - \vve B\ve z + \nabla(\vve B\ve z) \cdot
(\vve B\ve z))\abs{J^{-1}}\ \cdot\ \rho_w(\ve z)
	\end{eqnarray*}
because 
	$\rho_{xw}(\ve a, \ve b) = \rho_x(\ve a) \cdot \rho_w(\ve b)
 	= \rho(\ve a) \cdot \rho_w(\ve b)$ 
due to the independence between $\ve x$ and $\delw$. 
Since $S_y$ and $S_z$ are arbitrarily chosen, the integrand is the
very joint pdf $\rho_{yz}(\ve y, \ve z)$.
Thus
	\begin{eqnarray}
	&&
	\FP\rho(\ve y) = \rho_y (\ve y) 
		= \int_{\R^m} \rho_{yz}(\ve y, \ve z) d\ve z	\cr
	&&\qquad
	= \int_{\R^m} 
	  \bracket{\rho(\ve y - \ve F\delt - \vve B\ve z 
		   + \nabla(\vve B\ve z) \cdot (\vve B\ve z))
		     \cdot \abs{J^{-1}}
		  } \ \cdot\ \rho_w(\ve z)\ d\ve z	\cr
	&&\qquad
	= E_w \cbrace{\rho(\ve y - \ve F\delt - \vve B \delw 
		   + \nabla(\vve B\delw) \cdot (\vve B \delw))
		     \cdot \abs{J^{-1}}		
		    } 					\cr
	&&\qquad
	= E_w 
		\bracket{\rho(\ve y) - \nabla\rho \cdot
			(\ve F\delt + \vve B\delw + 
			 \nabla(\vve B\delw) \cdot (\vve B\delw)) +
			 \frac12 (\vve B\delw) (\vve B\delw)^T :
			 \nabla\nabla\rho 
			}		\cr
	&&\qquad\qquad
		     \cdot 
		\bracket{1 - \nabla\cdot\ve F\delt - 
			\nabla\cdot(\vve B\delw) + 
			\frac12 \nabla\nabla : 
			(\vve B\delw \delw^T \vve B^T) }
		+ \hot		\cr
	&&\qquad
	= \rho(\ve y) - (\ve F\cdot \nabla\rho + \rho\nabla\cdot\ve F)\delt \cr
	&&\qquad\qquad
	  + \frac12 \bracket{ \rho \nabla\nabla : (\vve B\vve B^T)
		+ 2 \nabla\rho \cdot [\nabla\cdot(\vve B\vve B^T)]
		+ (\vve B\vve B^T) : (\nabla\nabla\rho)} \delt	
		+ \hot \cr
	&&\qquad
	= \rho(\ve y) - \nabla\cdot(\ve F\rho) \delt 
		      + \frac12 \nabla\nabla : (\vve B\vve B^T\rho)\delt
		+ \hot.
	\end{eqnarray}
Note here the fact
	\begin{eqnarray}
	E\delw = \ve 0,\qquad
	E\delw \delw^T = \delt \vve I
	\end{eqnarray}
about Wiener process has been used.
As a verification, one may obtain from this step
	\begin{eqnarray}
	\Dt\rho = \lim_{\delt\to0} \frac {\FP\rho(\ve y) - \rho(\ve y)} \delt
	= -\nabla\cdot(\ve F\rho) + \frac12 \nabla\nabla : 
			(\vve B\vve B^T\rho)
	\end{eqnarray}
which is precisely the Fokker-Planck equation (cf. the appendix).

Denote $\vve B\vve B^T$ by $\vve G$. Integrate both sides of 
the above equation with respect to $(y_2,y_3,...,y_n)$ to obtain
	\begin{eqnarray*}
	(\FP\rho)_1(y_1) 
	= \rho_1(y_1) - 
	  \delt\int_{\R^{n-1}} \DyI {F_1\rho} dy_2...dy_n +
	  \frac\delt 2 \int_{\R^{n-1}} \DyIDyI {g_{11}\rho} dy_2...dy_n
	  + \hot,
	\end{eqnarray*}
and hence
	\begin{eqnarray}	\label{eq:stoch_cont_tmp1}
	\log(\FP\rho)_1(y_1) = \log\rho_1(y_1) -
	  \frac\delt {\rho_1} \int_{\R^{n-1}} \DyI {F_1\rho} dy_2...dy_n +
	  \frac\delt {2\rho_1} 
			\int_{\R^{n-1}} \DyIDyI {g_{11}\rho} dy_2...dy_n
	  + \hot.
	\end{eqnarray}
So 
	\begin{eqnarray*}
	H_1(t+\delt) = -E\log(\FP\rho)_1(y_1) = -E\log\rho_1(y_1)
	\end{eqnarray*}
as the rest two terms vanish after applying the operator
$E(\cdot) = \int_\R \rho_1 (\cdot) dy_1$.
Expanding $y_1$ around $x_1$, and denoting 
$\ve B_1 \equiv (b_{11},b_{12},...,b_{1n})$,
we have
	\begin{eqnarray*}
	&&
	H_1(t+\delt) = -E\log\rho_1(x_1 + F_1\delt + \ve B_1\delw) \cr
	&&\qquad
	= -E\bracket{
    	    \log\rho_1(x_1) + \DI {\log\rho_1} (F_1\delt + \ve B_1\delw)
		+ \frac12 \DIDI {\log\rho_1} \ve B_1\delw \delw^T \ve B_1^T
		    }	+ \hot \cr
	&&\qquad
	= H_1(t) - E\bracket{F_1 \DI {\log\rho_1}} \delt
		  - \frac12 E\bracket{g_{11} \DIDI {\log\rho_1}} \delt
		    + \hot.
	\end{eqnarray*}
Let $\delt\to 0$ and we finally arrive at
	\begin{eqnarray*}
	\dt {H_1} = -E\bracket{F_1 \DI {\log\rho_1}}
		  - \frac12 E\bracket{g_{11} \DIDI {\log\rho_1}}.
	\end{eqnarray*}
\qed

Now consider during the time interval $[t,t+\delt]$ to freeze $x_2$ as a
parameter, and examine how the marginal entropy of $x_1$ evolves.
In this case we are actually considering a density $\rho_{1\excl2}$, with
$rho_{1\excl2}(t) = \rho_1(t)$ under an $(n-1)$-dimensional transformation:
$\R^{n-1} \to \R^{n-1}$, $\ve x_{1\excl2} \to \ve y_{1\excl2}$:
	\begin{eqnarray*}
	\left\{\begin{array}{l}
	y_1 = x_1(t+\delt) = x_1(t) + F_1\delt + \ve B_1\delw, \\
	y_3 = x_3(t+\delt) = x_3(t) + F_3\delt + \ve B_3\delw, \\
	... \\
	y_n = x_n(t+\delt) = x_n(t) + F_n\delt + \ve B_n\delw.	
	\end{array}\right.
	\end{eqnarray*}
With this system we have the following proposition.
   \begin{prop}
    Let $\rho_\excl2$ be $\int_\R \rho(\ve x) dx_2$, then
	\begin{eqnarray}	\label{eq:dH12}
   	\dt {H_{1\excl2}} 
	&=&
        -E\bracket{F_1\DI{\log\rho_1}}
	- \frac12 E\bracket{g_{11}\DIDI {\log\rho_1} }	\cr
	&\ &
	+ E\bracket{\frac 1 {\rho_1} 
		\int_{\R^{n-2}} \DI {F_1\rho_\excl2} dx_3...dx_n  
		   }					\cr
	&\ &
	- \frac12 E\bracket{
		\frac 1 {\rho_1}
		\int_{\R^{n-2}} \DIDI {g_{11}\rho_\excl2} dx_3...dx_n 
			   }.
	\end{eqnarray}
   \end{prop}
\pf
    Following the same procedure as above, we arrive at an equation for
    $\log(\FP_\excl2)_1(y_1)$ similar to (\ref{eq:stoch_cont_tmp1}):
	\begin{eqnarray*}
	\log(\FP_\excl2\rho)_1(y_1) = \log\rho_{1\excl2}(y_1) -
	  \frac\delt {\rho_{1\excl2}} 
		\int_{\R^{n-2}} \DyI {F_1\rho_\excl2} dy_3...dy_n +
	  \frac\delt {2\rho_{1\excl2}} 
		\int_{\R^{n-2}} \DyIDyI {g_{11}\rho_\excl2} dy_3...dy_n
	  + \hot.
	\end{eqnarray*}
So 
	\begin{eqnarray*}
	&&
	H_{1\excl2}(t+\delt) = -E\log(\FP_\excl2\rho)_1(y_1)	\cr
	&&\qquad
	= -E\log\rho_{1\excl2}(y_1)  	\cr
	&&\qquad\ \ \
	 + E\bracket{\frac 1 {\rho_{1\excl2}} \int_{\R^{n-2}}
		\DyI{F_1\rho_\excl2} dy_3...dy_n} \delt \cr
	&&\qquad\ \ \
	 - \frac12 E\bracket{\frac 1 {\rho_{1\excl2}} \int_{\R^{n-2}}
		\DyIDyI{g_{11} \rho_\excl2} dy_3...dy_n} \delt
	 + \hot.
	\end{eqnarray*}
Note at time $t$, $\rho_{1\excl2} = \rho_1$, and 
in the last two terms $\ve y$ can be replaced by $\ve x$ with error going
to higher order terms. Thus
	\begin{eqnarray*}
	&&
	H_{1\excl2}(t+\delt) 
	= -E\bracket{
	      \log\rho_1(x_1) + \DI{\log\rho_1}(F_1\delt + \ve B_1\delw)
	      + \frac12 \DIDI {\log\rho_1} \ve B_1 \delw \delw^T \ve B_1^T
		    }	\cr
	&&\qquad\ \ \
	 + E\bracket{\frac 1 {\rho_1} \int_{\R^{n-2}}
		\DI{F_1\rho_\excl2} dx_3...dx_n} \delt \cr
	&&\qquad\ \ \
	 - \frac12 E\bracket{\frac 1 {\rho_1} \int_{\R^{n-2}}
		\DIDI{g_{11} \rho_\excl2} dx_3...dx_n} \delt
	 + \hot	\cr
	&&\qquad
	= H_1(t) 
        -E\bracket{F_1\DI{\log\rho_1}} \delt
	- \frac12 E\bracket{g_{11}\DIDI {\log\rho_1} }\delt	\cr
	&&\qquad\ \ \
	+ E\bracket{\frac 1 {\rho_1} 
		\int_{\R^{n-2}} \DI {F_1\rho_\excl2} dx_3...dx_n  
		   }  \delt					\cr
	&&\qquad\ \ \
	- \frac12 E\bracket{
		\frac 1 {\rho_1}
		\int_{\R^{n-2}} \DIDI {g_{11}\rho_\excl2} dx_3...dx_n 
			   } \delt 
	+ \hot.
	\end{eqnarray*}
Take the limit 
	$$\dt {H_{1\excl2}} = \lim_{\delt\to0} 
		\frac {H_{1\excl2}(t+\delt) - H_1(t)} 
			\delt$$
and we arrive at the conclusion.
\qed

   \begin{thm}
    	\begin{eqnarray}
	T_{2\to1} 
	&=& 
	- E\bracket{\frac 1 {\rho_1} 
		\int_{\R^{n-2}} \DI {F_1\rho_\excl2} dx_3...dx_n  
		   }
	+ \frac12 E\bracket{
		\frac 1 {\rho_1}
		\int_{\R^{n-2}} \DIDI {g_{11}\rho_\excl2} dx_3...dx_n 
			   }.	\\
	&=&
	- \int_{\R^n} \rho_{2|1} (x_2|x_1) \DI {F_1\rho_\excl2} d\ve x
	+ \frac12 \int_{\R^n} \rho_{2|1} (x_2|x_1) 
		\DIDI {g_{11}\rho_\excl2} d\ve x.
	\end{eqnarray}
   \end{thm}
\pf
Subtract (\ref{eq:dH12}) from (\ref{eq:dH1}) and the conclusion follows.
\qed


\subsection{Properties}
    \begin{thm}
    For a 2D system 
	\begin{eqnarray}
	&& \dt {H_{1\excl2}} = E\parenth{\DI {F_1}} \\
	\end{eqnarray}
    in the absence of stochasticity.
    \end{thm}
    \noindent Remark: This 
	recovers the heuristic argument by Liang and Kleeman in \cite{LK05};
	see Eq.~(\ref{eq:phys_argu}).

    \pf In this case $g_{11}=0$, $\rho_\excl2 = \rho_1$, so
	\begin{eqnarray*}
	\dt {H_{1\excl2} }
	&=& - E\bracket{F_1 \DI {\log\rho_1}}
			   + E\bracket{\frac 1 {\rho_1} \DI{F_1\rho_1}} \cr
	&=& E\bracket{\frac {\rho_1} {\rho_1} \DI{F_1} + F_1\DI{\log\rho_1}
			- F_1 \DI {\log\rho_1} } \cr
	&=& E\parenth{\DI {F_1}}.
	\end{eqnarray*}
    \qed


    \begin{thm}	\label{thm:stoch_vanishes}
     If $g_{11} = \sum_{k=1}^m b_{1k} b_{1k}$ is independent of $x_2$,
     the resulting $T_{2\to1}$ has a form same as its deterministic
     counterpart.
    \end{thm}
     \pf
     If $g_{11}$ is independent of $x_2$, so is
		$\int \DIDI {g_{11} \rho_\excl2} dx_3...dx_n.$
     Hence the integration can be simplified:
	\begin{eqnarray*}
	&&
	\int_{\R^n} \rho_{2|1} \DIDI {g_{11}\rho_\excl2} d\ve x
	= \int_{\R^{n-1}} 
	  \parenth{\int_\R \frac {\rho_{12}} {\rho_1} dx_2} 
	  \cdot
	  \DIDI {g_{11}\rho_\excl2} dx_1dx_3...dx_n \cr
	&&\qquad\qquad
	= \int_{\R^{n-1}} \DIDI {g_{11}\rho_\excl2} dx_1dx_3...dx_n 
	= 0.
	\end{eqnarray*}
    \qed

    \begin{thm} {\bf (Property of causality)}\\
     If both $F_1$ and $g_{11}$ are independent of $x_2$, 
     then $T_{2\to1} = 0$.
    \end{thm}
    \pf
     As proved above, when $g_{11}$ has no dependence on $x_2$, the last
     term of $T_{2\to1}$ becomes zero. If, moreover, $F_1$ does not depend
     on $x_2$, then $\DI {F_1\rho_\excl2}$ does not, either. So the
integration with respect to $x_2$ can be taken inside directly to
$\rho_{12}/\rho_1 = \rho_{2|1}(x_2 | x_1)$:
	\begin{eqnarray*}
	T_{2\to1} = -\int_\R dx_1 \int_\R \rho_{2|1} (x_2|x_1) dx_2 \cdot
		\int_{\R^{n-2}} \DI {F_1\rho_\excl2} dx_3...dx_n
	= \int_{\R^{n-2}} \DI {F_1\rho_\excl2} dx_3...dx_n = 0.
	\end{eqnarray*}
\qed

\subsection{Application: A stochastic gradient system}


We are about to study the information flow within a system which has a
drift function in the gradient form. We particularly want to understand how
stochastic perturbation may exert influence on the flow.
The gradient systems are chosen because their corresponding 
Fokker-Planck equation admit explicit equilibrium solutions, 
i.e., solutions of the Boltzmann type.  To see this, let 
	\begin{eqnarray}
	\ve F = - \nabla V,
	\end{eqnarray}
where $V = V(\ve x)$ is the potential function. For simplicity, suppose
that the stochastic perturbation amplitude $\vve B = b\vve I$ where $I$ is
the identity matrix and $b=\const$. Hence 
	$\vve G = \vve B \vve B^T = g \vve I$, and $g=b^2$ is a constant. 
It is trivial to verify that 
	\begin{eqnarray}
	\rho = \frac 1 Z e^{-2V/g},
	\end{eqnarray}
where $Z$ is the normalizer (or partition function as is called in
statistical physics), solves
	\begin{eqnarray*}
	\nabla\cdot(\rho\ve F) = \frac12 g \nabla^2\rho,
	\end{eqnarray*}
the equilibrium density equation for the system
	\begin{eqnarray}
	d\ve x = -\nabla V dt + b\vve I d\ve w.
	\end{eqnarray}

As an example, consider the potential function
	\begin{eqnarray}	\label{eq:pot_func}
	V = \frac12 (x_1^2 x_2^2 + x_2^2 x_3^2 + x_1^2 + x_2^2 + x_3^2).
	\end{eqnarray}
This system, though simple, results in a compactly supported density
function, while allowing for asymmetric nonlinear interactions among $x_1$,
$x_2$, and $x_3$. The resulting vector field is
	\begin{eqnarray*}
	&& F_1 = -x_1 x_2^2 - x_1,	\\
	&& F_2 = -x_2 x_3^2 - x_2 x_1^2 - x_2, \\
	&& F_3 = -x_3 x_2^2 - x_3.
	\end{eqnarray*}
Obviously, $T_{3\to1} = T_{1\to3} = 0$ by the theorem on causality property. 
The general flow from $x_j$ to $x_i$ is
	\begin{eqnarray}
	T_{j\to i} 
	&=& -\int_{\R^3} \rho_{j|i}(x_j | x_i)
	     \frac {\D F_i \rho_{\excl j}} {\D x_i}\ d\ve x  \cr
	&=& -\int_{\R^3} \rho_{j|i} 
	     \parenth{F_i \Di {\rho_{\excl j}} + 
		     \rho_{\excl j} \Di {F_i}} \cr
	&=& -\int_{\R^3} \rho_{j|i} \parenth{
		F_i \int_\R \frac2g \rho F_i dx_j + 
		\rho_{\excl j} \Di {F_i} } \ d\ve x.
	\end{eqnarray}
The computation seems to be easy, but by no means trivial. The difficulty
comes from the evaluation of the conditional density $\rho_{j|i}(x_j | x_i)$.
Theoretically this is not a problem, but in realizing the computation
we have to consider the problem on a limited domain, which may not
effectively cover the support of the density function. Here we choose
a domain $[-5,5] \times [-5,5] \times [-5,5]$, and a spacing size
$\Delta x = 0.05$. The computation is implemented henceforth.
   \begin{figure} [h]
   \begin{center}
   \includegraphics[angle=0,width=1\textwidth]
     {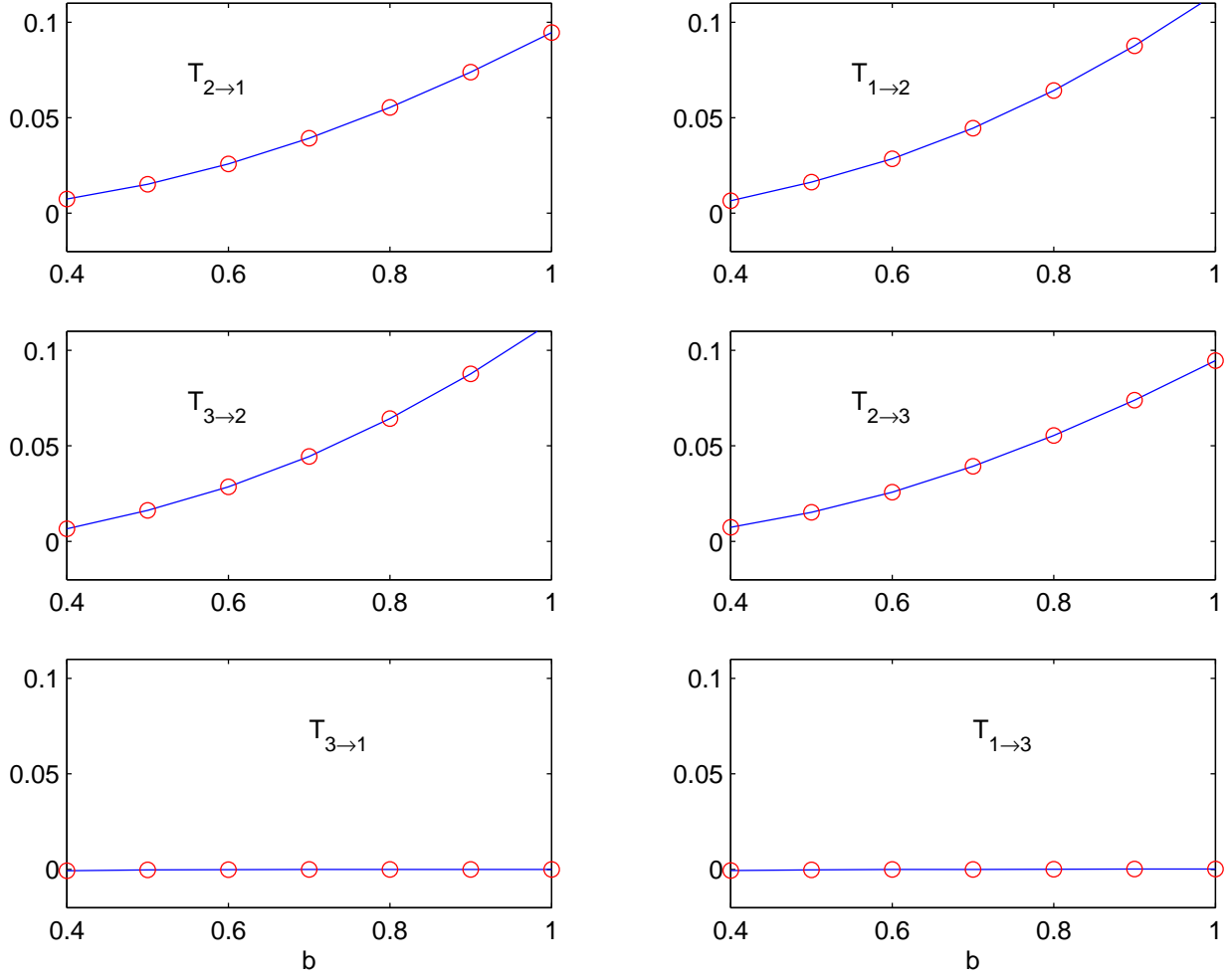}
   \caption{Information flow within a gradient system with the potential
	function (\ref{eq:pot_func}).}
        \protect{\label{fig:grad}}
   \end{center}
   \end{figure}

To test how the stochastic perturbation may affect the information flow.
tune $b$ to see the response. The tuning range is rather limited, though,
with the present computational domain. Shown in Fig.~\ref{fig:grad} are
the results. As expected, $T_{3\to1}$ and $T_{1\to3}$ are identically zero. 
For others, the flow rates generally increase with $b$. That is to say,
they tend to increase the uncertainty of their corresponding target
components. This makes sense, since $g$ functions like temperature in
thermodynamics, and increase in $T$ surely will lead to increase in
uncertainty. If examining more carefully, one finds that the increase is
actually not symmetric.  Those going to $x_2$ ($T_{3\to2}$ and $T_{1\to2}$)
are faster than those leaving $x_2$ ($T_{2\to1}$ and $T_{2\to3}$),
reflecting the property of asymmetry of information flow.

Since $\rho$ can be accurately
obtained, this example can be utilized to validate our numerical
computations for more general cases.

\section{Linear stochastic systems}	\label{sect:linear}
As always, it would be of interest to look at the particular case, namely,
the case with linear systems:
	\begin{eqnarray}
	d\ve x = \vve A \ve x dt + \vve B d\ve w,
	\end{eqnarray}
with $\vve A$ and $\vve B$ being constant matrices.
If originally $\ve x$ is normally distributed, then it is normal/Gaussian
forever. Let its mean vector be $\veg\mu$ and its covariance matrix be
$\vveg\Sigma$. Then
	\begin{eqnarray}
	&&\dt{\veg\mu} = \vve A \veg\mu,	\\
	&&\dt{\vveg\Sigma} = \vve A \vveg\Sigma + \vveg\Sigma \vve A^T
		+ \vve B \vve B^T.	
	\end{eqnarray}
In component form $\veg\mu = (\mu_1,...,\mu_n)^T$,
$\vveg\Sigma = \parenth{\sigma_{ij}}_{n\times n}$,
and $\vve B \vve B^T$ has been denoted by $\vve G$ in the above.
The distribution is, therefore,
	\begin{eqnarray*}
	\rho = \frac 1 {\sqrt{(2\pi)^n \det \vveg\Sigma }}
	 e^{-\frac12 (\ve x - \veg\mu)^T \vveg\Sigma^{-1} (\ve x - \veg\mu)}.
	\end{eqnarray*}
We need to find
	$$\rho_1, \rho_{12}, \rho_\excl2$$, 
and the following facts will help.

Fact 1:
    $\rho_\excl2$ is a multivariate Gaussian 
	$N(\veg\mu_\excl2, \vveg\Sigma_\excl2)$
    where $\veg\mu_\excl2 = (\mu_1, \mu_3, \mu_4, ..., \mu_n)^n$,
    and $\vveg\Sigma_\excl2$ is the covariance matrix of
          $(x_1, x_3, x_4, ..., x_n)^n$.

Fact 2:
   The conditional probability density function $\rho_{2|1}$ is
	\begin{eqnarray}
	\rho_{2|1} (x_2 | x_1) \propto
		e^{-\frac {\sigma_{11}} {2\Delta_{12}} 
		    \bracket{x_2 - \mu_2 - \frac{\sigma_{12}} {\sigma_{11}}
		     (x_1 - \mu_1)}^2},
	\end{eqnarray}
   in other words, 
	\begin{eqnarray}
	x_2 | x_1 \sim 
	N\parenth{\mu_2 + \frac{\sigma_{12}} {\sigma_{11}} (x_1 - \mu_1),\
		 \frac {\Delta_{12}} {\sigma_{11}}}.
	\end{eqnarray}
In the above equations, we have used, and will be using, $\Delta_{ij}$ to 
shorten 
	$\det \mat {\sigma_{ii}}  {\sigma_{ij}}  
		   {\sigma_{ij}} {\sigma_{jj}}.$

We now compute the information flow $T_{2\to1}$. Since 
$\vve B$ is constant (hence independent of $x_1$), the stochastic term
vanishes by Theorem~\ref{thm:stoch_vanishes}. So we need only consider 
its deterministic part:
	\begin{eqnarray*}
	T_{2\to1} = -E\bracket{\frac1{\rho_1} \int_{\R^{n-2}} 
		    \DI {F_1\rho_\excl2}}
	= - \int_{\R^n} \rho_{2|1} (x_2 | x_1) \DI {F_1\rho_\excl2} d\ve x.
	\end{eqnarray*}
As a starting point, let us consider the case $n=3$. 
By the proposition above,
	\begin{eqnarray*}
	\rho_\excl2 = \rho_{13} 
	= \frac 1 {\sqrt{(2\pi)^2 \Delta_{13}}}
	   e^{- \frac 1 {\Delta_{13}} [\sigma_{33} (x_1-\mu_1)^2 +
	 \sigma_{11} (x_3-\mu_3)^2 ] - 2\sigma_{13}(x_1-\mu_1)(x_3-\mu_3)]}.
	\end{eqnarray*}
So
	\begin{eqnarray*}
	&& \int_\R \DI {F_1 \rho_\excl2} dx_3
	   = \int_\R \rho_{13} \cbrace{ a_{11}   +   
	   [\sigma_{13} (x_3 - \mu_3) - \sigma_{33} (x_1 - \mu_1)]
	   \cdot (a_{11}x_1 + a_{12} x_2 + a_{13} x_3) / \Delta_{13}
				} dx_3	\\
	&& =
	   a_{11} \rho_1 - \frac {\sigma_{13}\mu_3 +
			\sigma_{33}(x_1-\mu_1)(a_{11}x_1 + a_{12}x_2)} 
			         {\Delta_{13}} \rho_1	\\
	&&\ \ \
	  + \frac 1 {\Delta_{13}} \int_\R \rho_{13} \cdot
		\bracket{
			a_{13}\sigma_{13} x_3^2 + 
			(a_{11}x_1 + a_{12}x_2) \sigma_{13}x_3 - 
		        (\sigma_{13}\mu_3 + \sigma_{33}(x_1 - \mu_1)) 
				a_{13} x_3 
		       } dx_3.
	\end{eqnarray*}
We need to find $\int_\R x_3 \rho_{13} dx_3$ and 
		$\int_\R x_3^2 \rho_{13} dx_3$.
Since $(x_1,x_3)$ is a bivariate Gaussian, 
	\begin{eqnarray*}
	x_3 | x_1 \sim 
	N\parenth{\mu_3 + \frac{\sigma_{13}} {\sigma_{11}} (x_1 - \mu_1),\
		 \frac {\Delta_{13}} {\sigma_{11}}},
	\end{eqnarray*}
we thence have
	\begin{eqnarray*}
	&& \int_\R \rho_{13} x_3 dx_3 = \rho_1 \int_\R \rho_{3|1} x_3 dx_3
	   = \rho_1 \cdot \parenth{\mu_3 + 
	     \frac {\sigma_{13}}{\sigma_{11}} (x_1-\mu_1)},	\\
	&& \int_\R \rho_{13} x_3^2 dx_3 = \rho_1 \int_\R \rho_{3|1} x_3^2 dx_3
	   = \rho_1 \cdot \bracket{ \frac {\Delta_{13}} {\sigma_{11}}
	+ \parenth{\mu_3 + \frac{\sigma_{13}}{\sigma_{11}} (x_1 - \mu_1)}^2
				  }.
	\end{eqnarray*}
Substituting back, we obtain:
	\begin{eqnarray*}
	&& \int_\R \DI {F_1\rho_\excl2} dx_3 =
	   a_{11} \rho_1 - \frac {\sigma_{13}\mu_3 +
			\sigma_{33}(x_1-\mu_1)(a_{11}x_1 + a_{12}x_2)} 
			         {\Delta_{13}} \rho_1	\\
	&&\qquad
	   + a_{13}\sigma_{13} \parenth{
		\frac{\Delta_{13}} {\sigma_{11}}
        + \bracket{\mu_3 + \frac{\sigma_{13}} {\sigma_{11}} (x_1-\mu_1)}^2
		   		      } \frac {\rho_1} {\Delta_{13}} \\
	&&\qquad
	 + \bracket{(a_{11}x_1 + a_{12}x_2) \sigma_{13} 
		- (\sigma_{13}\mu_3 + \sigma_{33}(x_1-\mu_1)) a_{13}}
	   \bracket{\mu_3 + \frac{\sigma_{13}}{\sigma_{11}} (x_1-\mu_1)}
	   \frac {\rho_1} {\Delta_{13}}.
	\end{eqnarray*}
Thus
	\begin{eqnarray*}
	T_{2\to1} 
	&=& - E\frac 1 {\rho_1} \DI{F_1\rho_{13}} dx_3	\\
 	&=& -a_{11} 
	     - \frac 1 {\Delta_{13}} 
	 	[-\sigma_{13}\mu_3 a_{11} \mu_1 
		 - \sigma_{13}\mu_3 a_{12}\mu_2
	 	 - a_{11} \sigma_{33} \sigma_{11}	
		 - a_{12} \sigma_{33} \sigma_{12}	\\
	&&\qquad\
		 + a_{13} \sigma_{13} \frac {\Delta_{13}} {\sigma_{11}}
		 + a_{13} \sigma_{13} 
	     	   (\mu_3^2 + \sigma_{13}^2 / \sigma_{11}^2 \cdot \sigma_{11})
		 + a_{11} \sigma_{13} \mu_3 \mu_1	\\
	&&\qquad\
	  	 + a_{12} \mu_3 \sigma_{13} \mu_2
		 - a_{13} \sigma_{13} \mu_3^2 
		 - 0
	 	 + a_{11} \sigma_{13}^2/\sigma_{11} \cdot \sigma_{11} \\
	&&\qquad\
		 + a_{12} \sigma_{13}^2 / \sigma_{11} \cdot \sigma_{12}
		 - 0					
		 - a_{13} \sigma_{33} \sigma_{13} / \sigma_{11}
		   \cdot\sigma_{11}]	\\
	&=& a_{12} \frac {\sigma_{12}} {\sigma_{11}}.
	\end{eqnarray*}
The so many terms are canceled out, and the result turn out to be
precisely the same as that for the 2D case we have derived before ever
since Liang and Kleeman (2005)!

The above remarkably concise formula actually holds for systems of 
arbitrary dimensionality. This makes the following theorem:
	\begin{thm}	\label{thm:linearflow}
	If an $n$-dimensional ($n\ge2$) vector of random variables 
	$(x_1,...,x_n)^T$ evolves subject to the linear system
		\begin{eqnarray*}
		d\ve x = \vve A \ve x dt + \vve B d\ve w,
		\end{eqnarray*}
	where $\vve A = (a_{ij})$ and $\vve B$ are constant matrices, 
	and if its covariance matrix is $(\sigma_{ij})$, then
	the information flow from $x_j$ to $x_i$ is
		\begin{eqnarray}	\label{eq:linearflow}
		T_{j\to i} = a_{ij} \frac {\sigma_{ij}} {\sigma_{ii}},
		\end{eqnarray}	
	for any $i,j=1,...,n$, $i\ne j$.
	\end{thm}
\pf
It suffices to prove the case $(i,j) = (1,2)$; if not, we may always
reorder the components to make them so. 
We prove by induction. The 3D case has just been shown above.
Now suppose (\ref{eq:linearflow}) holds for $n$-dimensional systems.
Consider an $n$+1-dimensional system
	\begin{eqnarray*}
	&&\dt {x_1} = \sum_{j=1}^n a_{1j} x_j + a_{1,n+1} x_{n+1}, \\
	 && \quad \vdots	\qquad\qquad\qquad\qquad \vdots	  \\
	&&\dt {x_2} = \sum_{j=1}^n a_{nj} x_j + a_{n,n+1} x_{n+1} \\
	&&\dt {x_{n+1}} = \sum_{j=1}^n a_{n+1,j} x_j + a_{n+1,n+1} x_{n+1}.
	\end{eqnarray*}
To distinguish, we now use $\rho^n$ to denote the joint density for the
$n$-dimensional system.
The information flow from $x_2$ to $x_1$ is
	\begin{eqnarray*}
	T_{2\to1} 
	&=& -\int_{\R^{n+1}} \rho_{2|1} (x_2 | x_1) 
		     \DI{F_1\rho_\excl2} d\ve x	\\
	&=& \int_{\R^{n+1}} \rho_{2|1} 
	    \DI\ \bracket{(\sum_{j=1}^n a_{1j} x_j) \rho_\excl2 + 
		(a_{1,n+1} x_{n+1}) \rho_\excl2} d\ve x	\\
	&=&
	    \int_{\R^n} \rho_{2|1} \DI\ \parenth{\sum_{j=1}^n 
		a_{1j} x_j \rho_\excl2^n} 
	    + 
	    \int_{\R^{n+1}} \rho_{2|1} 
		\DI\ \parenth{a_{1,n+1} x_{n+1} \rho_\excl2} 
		d\ve x.
	\end{eqnarray*}
Note the first term results from integration with respect $x_{n+1}$,
since all the variables except $\rho_\excl2$ are independent of $x_{n+1}$.
This is precisely the information flow from $x_2$ to $x_1$ for an
$n$-dimensional system; by our assumption it is 
		$a_{12} \sigma_{12} / \sigma_{11}$.
For the second term, note that all variables, except $\rho_{2|1}$, are
independent of $x_2$, so we may take integral with respect to $x_2$
directly inside with $\rho_{2|1}$. But $\int_\R \rho_{2|1} dx_2 = 1$, 
so the second term results in the integral of
	$\DI\ (a_{1,n+1} x_{n+1} \rho_\excl2)$
which vanishes by the compactness of $\rho$. Therefore (\ref{eq:linearflow})
holds for $n$+1-dimensional systems. By induction, it holds for systems of
arbitrary dimensionality. \qed

\begin{figure} [h]
   \begin{center}
   \includegraphics[angle=0,width=1\textwidth]
     {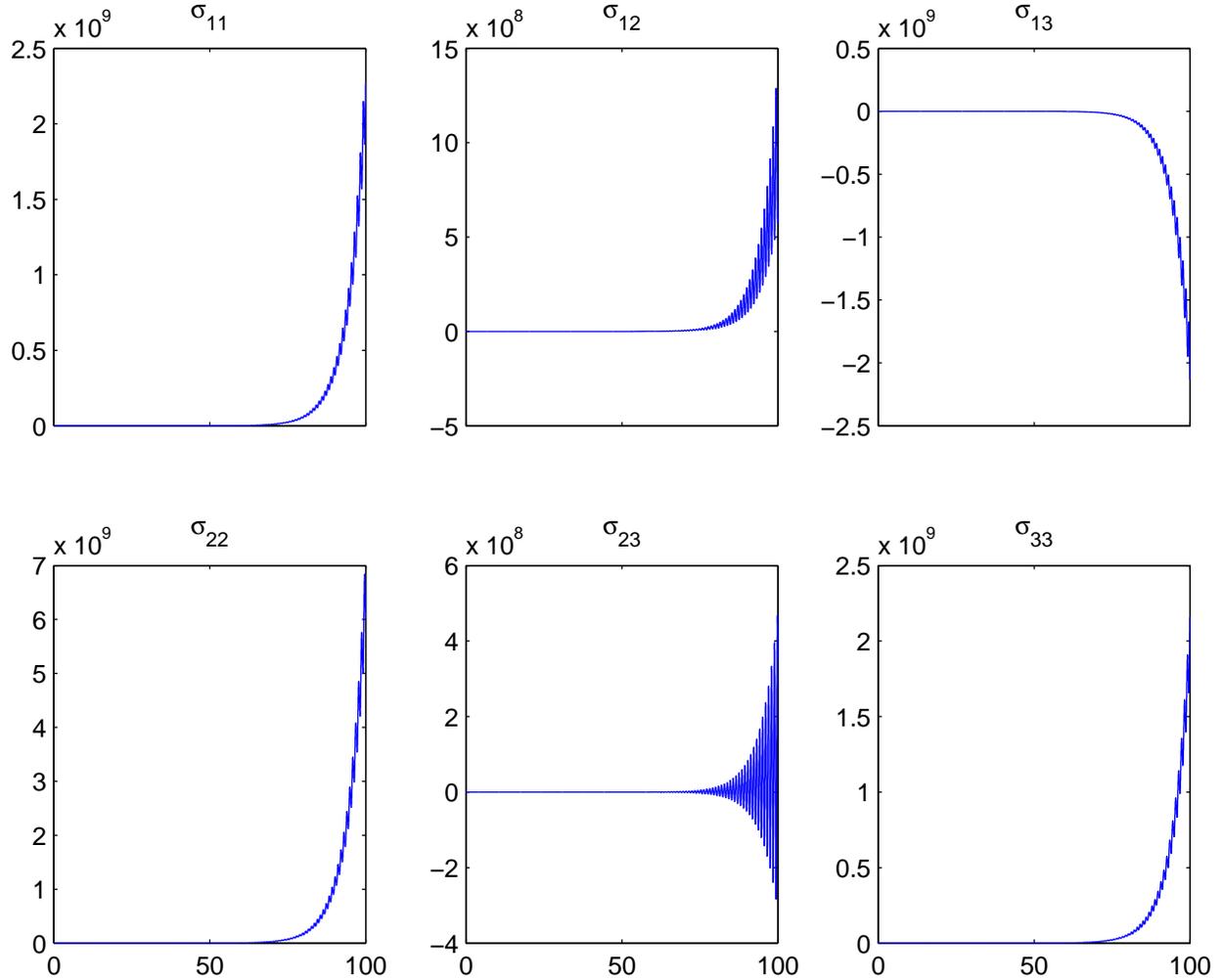}
   \caption{Covariance evolution with the linear system 
	(\ref{eq:linear1})-(\ref{eq:linear3}). 
		}
        \protect{\label{fig:linear_covariance}}
   \end{center}
   \end{figure}

\begin{figure} [h]
   \begin{center}
   \includegraphics[angle=0,width=1\textwidth]
     {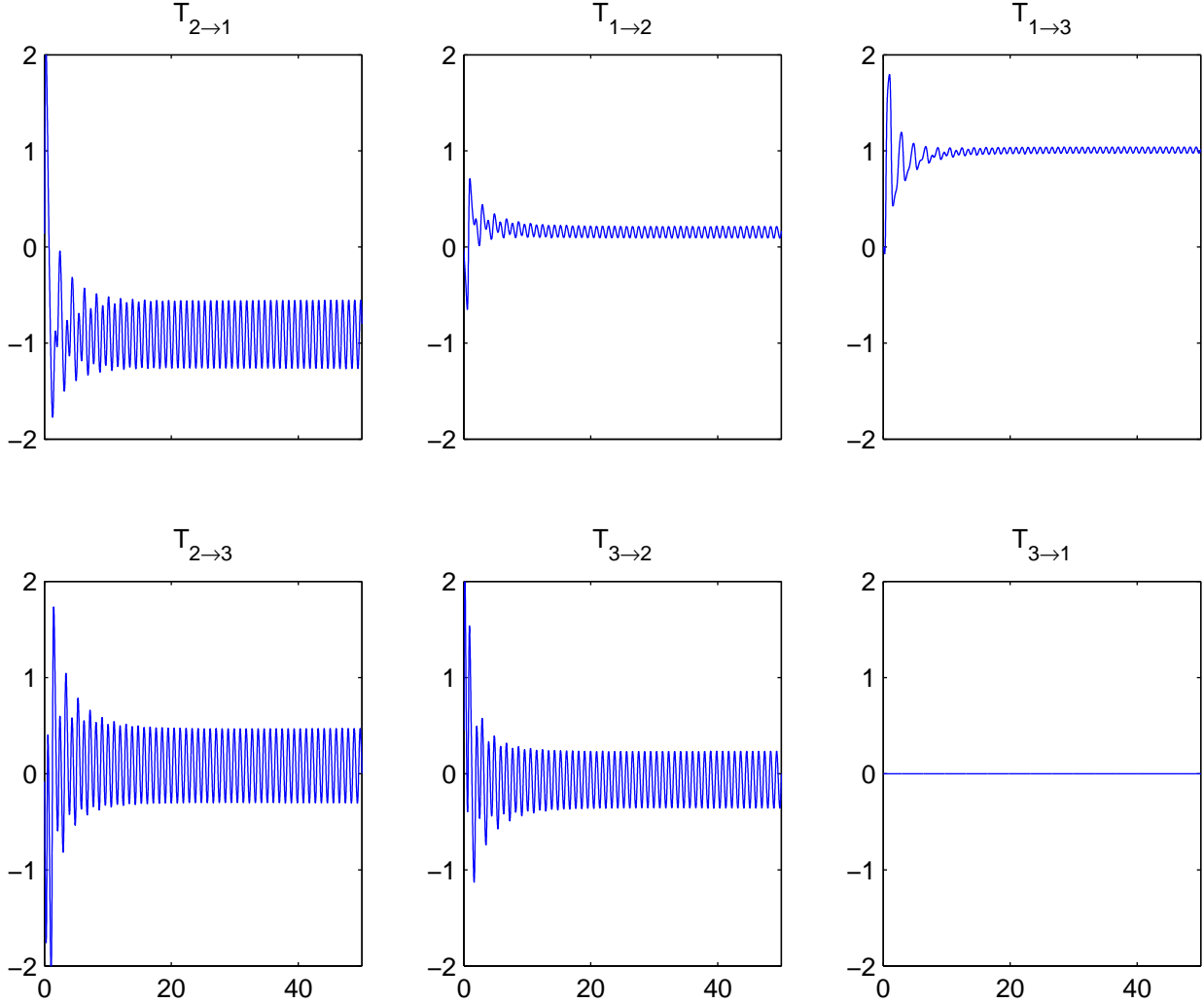}
   \caption{As Fig.~\ref{fig:linear_covariance}, but for rates 
	of information. }
        \protect{\label{fig:linear_T}}
   \end{center}
   \end{figure}

Let us see an example: $\vve A = \matthree  {1}  {-2}  {0}
					    {1}  {0}   {-5}
					    {-1} {2}   {-1}$, 
		  and  $\vve B = \matthree {1}  0     0
					    0  {2}   0
					    0    0   {3}.$
In component form, the equation is
	\begin{eqnarray}
	&&\dt {x_1} = x_1 - 2x_2 + 0 x_3 +  \dot w_1, \label{eq:linear1}\\
	&&\dt {x_2} = x_1 + 0 x_2 - 5 x_3 + 2 \dot w_2,\\
	&&\dt {x_3} = - x_1 + 2 x_2 - x_3 + 3 \dot w_3. \label{eq:linear3}
	\end{eqnarray}
The evolution of the covariance matrix $\vve C$ is governed by
	\begin{eqnarray}
	\dt {\vve C} = \vve A \vve C + \vve C \vve A^T + \vve B \vve B^T.
	\end{eqnarray}
Let it be initialized by $\matthree 1 0 0
				    0 4 0
				    0 0 9$.
The solution is shown in Fig.~\ref{fig:linear_covariance}.
The rates of information flow are subsequently obtained and plotted in 
Fig.~\ref{fig:linear_T}.
Among them, $T_{3\to1} = 0$, just as expected by the property of causality.
$T_{3\to2}$ and $T_{2\to3}$ oscillate around a value near zero, and
$T_{2\to1}$ oscillates around -0.9. The remaining transfers, 
$T_{1\to2}$ and $T_{1\to3}$, albeit still oscillatory, approximately
approach two constant values. The former approaches 0.16, 
while the latter approaches 1.


\section{Summary}	\label{sect:summary}

Information flow, or information transfer as it may appear in the
literature, is a fundamental notion in general physics which has wide
applications in different disciplines. In this study we have shown that,
within the framework of dynamical systems, it can be rigorously
derived from first principles. That is to say, it is a notion {\it ab
initio}, quite different from the existing axiomatic postulates 
or empirical proposals. 
In this light we have studied the information flow for
both time-discrete and time-continuous differentiable
vector fields in both deterministic and stochastic settings.
In a nutshell, the results can be summarized as follows. 
%
%

Consider an $n$-dimensional state variable $\ve x = (x_1,x_2,...x_n)$, the
corresponding probability density function (pdf) being $\rho(x_1,x_2,...x_n)$,
and the marginal pdf of $x_i$ being $\rho_i$.
For a deterministic mapping $\mapping: \R^n \to \R^n$, 
	\begin{eqnarray*}
    \ve x(\tau) \mapsto \ve x(\tau+1) = (\mapping_1(\ve x),
	\mapping_2(\ve x),...\mapping_n(\ve x)),
	\end{eqnarray*}
the rate of information flowing from $x_2$ to $x_1$ proves to be
	\begin{eqnarray*}
	T_{2\to1} = E\log(\FP_\excl2\rho)_1(\mapping_1(\ve x))
	   - E\log(\FP\rho)_1(\mapping(_1(\ve x)),
	\end{eqnarray*}
where $E$ is the mathematical expectation with respect to $\ve x$, 
$\FP$ the Frobenius-Perron operator of $\mapping$, and 
$\FP_\excl2$ the same operator of $\mapping$ but with $x_2$ frozen as a
parameter (so $(\FP_\excl2)_1(x_1)$ has dependence on $x_2$). The units are
in nats per unit time; same below.
If the system is continuous in time, i.e.,
	\begin{eqnarray*}
	\dt {\ve x} = \ve F(\ve x, t),
	\end{eqnarray*}
then
	\begin{eqnarray*}
	T_{2\to1} = -\int_{\R^n} \rho_{2|1} \DI {\rho_\excl2 F_1} d\ve x
   = -E\bracket{\frac1 {\rho_1} \int_{\R^{n-2}} \DI{\rho_\excl2 F_1}
	dx_3...dx_n},
	\end{eqnarray*}
where $\rho_\excl2 = \int_\R \rho(x_1,x_2,...,x_n) dx_2$,
and $\rho_{2|1}$ is the conditional pdf of $x_2$ on $x_1$.
When stochasticity comes in, in the discrete mapping case:
	\begin{eqnarray*}
	\ve x(\tau+1) = \mapping(\ve x(\tau))) + \vve B(\ve x) \ve w,
	\end{eqnarray*}
where $\mapping: \R^n \to \R^n$ is an $n$-dimensional mapping, $\vve B$ an
$n\times m$ constant matrix, 
and $\ve w$ an $m$-dimensional standard Wiener process, then
	\begin{eqnarray*}
	T_{2\to1} = E_x [\log E_w(\FP_{\mapping_\excl2}\rho)_1
	(y_1 - \ve B_1\ve w)] - E_x [\log E_w(\FP_\mapping\rho)_1
	(y_1 - \ve B_1 \ve w)],
	\end{eqnarray*}
with $\ve B_1 = (b_{11}, b_{12},...,b_{1m})$ a row vector of the matrix 
$\vve B$. Here we use $E_x$ and $E_w$ to indicate that the expectation is
taken with respect to $x$ and $w$, respectively.
If what we consider is a continuous-time stochastic system, i.e.,
a system as
	\begin{eqnarray*}
	d\ve x = \ve F(\ve x, t) dt + \vve Bd\ve w,
	\end{eqnarray*}
or alternatively written as 
	\begin{eqnarray*}
	\dt {\ve x} = \ve F(\ve x,t) + \vve B \dot {\ve w},
	\end{eqnarray*}
where $\dot{\ve w}$ is the white noise, then the result can be explicitly 
evaluated:
    	\begin{eqnarray}
	T_{2\to1} 
	&=& 
	- E\bracket{\frac 1 {\rho_1} 
		\int_{\R^{n-2}} \DI {F_1\rho_\excl2} dx_3...dx_n  
		   }
	+ \frac12 E\bracket{
		\frac 1 {\rho_1}
		\int_{\R^{n-2}} \DIDI {g_{11}\rho_\excl2} dx_3...dx_n 
			   }.	\\
	&=&
	- \int_{\R^n} \rho_{2|1} (x_2|x_1) \DI {F_1\rho_\excl2} d\ve x
	+ \frac12 \int_{\R^n} \rho_{2|1} (x_2|x_1) 
		\DIDI {g_{11}\rho_\excl2} d\ve x,
	\end{eqnarray}
where $g_{11} = \sum_{j=1}^m b_{1j} b_{1j}$.
Note the first term is just from the deterministic vector field, while the
second the contribution from the noise. It has been
proved that, if $b_{1j}$ has no dependence on $x_2$, then the  stochastic
contribution vanishes, making the information flow same in form as that
from its deterministic counterpart.
We have particularly examined the case $\ve F = \vve A \ve x$, i.e.,
the case when the system is linear and autonomous, 
	\begin{eqnarray*}
	d\ve x = \vve A \ve x dt + \vve B d\ve w
	\end{eqnarray*}
with $\vve A = (a_{ij})_{n\times n}$ and $\vve B = (b_{ij})_{n\times m}$ 
being constant matrices, 
then the information flow from $x_j$ to $x_i$ is remarkably simple:
	\begin{eqnarray*}
	T_{j\to i} = a_{ij} \frac {\sigma_{ij}} {\sigma_{ii}},
	\end{eqnarray*}
for any $(i,j)$, $1\le i,j\le n$, $i\ne j$.
This result is precisely the same in form as originally we obtained for 2D
deterministic systems based on intuitive arguments in \cite{LK05}. 


Historically it has been a long-time endeavor to relate information flow to
causality. We want specifically to have that,
if $T_{j\to i} \ne 0$, then $x_j$ causes $x_i$, 
otherwise $x_j$ is not causal.
With the existing empirical/half-empirical measures for 
information flow, such as the widely used transfer entropy, 
the endeavor has been fruitful for some problems but unsuccessful 
for others (e.g., \cite{Smirnov13}), and the inconsistency has even led 
to doubt about the association between information flow and causality
(e.g., \cite{Lizier}). 
In this study, the implied causality, the touchstone one-way causality
in particular, is a proved fact for dynamical systems, as
stated in various theorems. More specifically, when the evolution of $x_i$
does not depend on $x_j$, then $T_{j\to i} = 0$.
This is particularly clear in the above linear case, the dependence of
$x_i$ on $x_j$ is from the entry $a_{ij}$ of $\vve A$, so when it is zero,
then $x_j$ is not causal to $x_i$. This result also quantitatively, and
unambiguously, tells us that, causation implies correlation, but
not vice versa, resolving the long-standing debate
over correlation versus causation.

The above results have been put to applications with a variety of benchmark
systems. Particularly we have re-examined the baker transformation, \Henon\
map, and truncated Burgers-Hopf system. The results are qualitatively
similar to what we have obtained before using an approximate formalism, 
but with magnitudes significantly smaller. Also shown are 
the information flows within a
Kaplan-Yorke map, a noisy \Henon\ map, a R\"ossler system,
and a stochastic gradient flow. We look forward to more applications to
real world problems in the near future.

\section*{Acknowledgment}
This study was supported by the Jiangsu Provincial Government through the
``Specially Appointed Professor Program'' (Jiangsu Chair Professorship) to
X.S.L., and by the Ministry of Finance of China through the Basic Research
Funding to China Institute for Advanced Study.

\end{document}